\documentclass[10pt, conference, letterpaper]{IEEEtran}
\IEEEoverridecommandlockouts
\usepackage{caption}
\usepackage{url}
\usepackage[english]{babel}
\usepackage[utf8]{inputenc}
\usepackage{algorithm}
\usepackage[noend]{algpseudocode}
\usepackage{subfiles}
\usepackage{amsmath}
\usepackage{amssymb}
\usepackage{mathrsfs}
\usepackage{amsthm,amsfonts}
\usepackage{bm}
\usepackage{bbm}
\usepackage{color}
\usepackage{subfigure}
\usepackage{xspace}
\newcommand{\eat}[1]{}
\usepackage[dvipsnames]{xcolor}

\usepackage{graphicx}

\newtheorem{theorem}{Theorem}
\newtheorem{definition}{Definition}

\newcommand{\diffperf}{\texttt{DiffPerf}\xspace}

\begin{document}

\title{DiffPerf: Towards Performance Differentiation and Optimization with SDN Implementation}
 \author{Walid Aljoby$^\dag$, Xin Wang$^\dag$, Dinil Mon Divakaran$^*$, Tom Z. J. Fu$^\S$, Richard T. B. Ma$^\dag$\\
 $^\dag$ School of Computing, National University of Singapore\\
 $^*$ Trustwave, Singapore\qquad $^\S$ Bigo Technology Pte Ltd, Singapore\\
 {\it \{algobi, xin.wang, tbma\}@comp.nus.edu.sg, dinil.divakaran@trustwave.com, fuzhengjia@gmail.com}
 }
\maketitle
\IEEEpeerreviewmaketitle


\begin{abstract}

Continuing the current trend, Internet traffic is expected to grow significantly over the coming years, with video traffic consuming the biggest share. On the one hand, this growth poses challenges to access providers (APs), who have to upgrade their infrastructure to meet the growing traffic demands as well as find new ways to monetize their network resources. On the other hand, despite numerous optimizations of the underlying transport protocol, a user's utilization of network bandwidth and is thus the user's perceived quality still being largely affected by network latency and buffer size. To address both concerns, we propose DiffPerf, a class-based differentiation framework, that, at a macroscopic level dynamically allocates bandwidth to service classes pre-defined by the APs, and at a microscopic level statistically differentiates and isolates user flows to help them achieve better performance. We implement DiffPerf on OpenDaylight SDN controller and programmable Barefoot Tofino switch and evaluate it from an application perspective for MPEG-DASH video streaming. Our evaluations demonstrate the practicality and flexibility that DiffPerf provides APs with capabilities through which a spectrum of qualities are provisioned at multiple classes. Meanwhile, it assists in achieving better fairness and improving overall user's perceived quality within the same class.



\end{abstract}

 \section{Introduction}

Today's Internet is dominated by content traffic, especially video streams. 
According to Cisco Annual Internet Report~\cite{ciscoReport}, video will make up 82\% of the total downstream Internet traffic by 2022. In today's home, Internet video drives our work and life, particularly with COVID-19 pandemic~\cite{CANOPUS}, meanwhile video applications continue to be of a significant demand for the bandwidth in the future~\cite{ciscoReport}.
To accommodate high traffic, content providers (CPs) have been deploying wide-area infrastructures to bring content closer to users, e.g., Netflix uses third-party content delivery networks such as Akamai and Limelight, and builds its own \cite{OpenConnect}. 
However, as end-users rely on last-mile access providers (APs) for accessing the Internet, APs' bandwidth capacity still limit user throughput due to network congestion~\cite{kundelqueueing}. For example, the average throughput of Netflix users behind Comcast~\cite{Netflix}, the largest U.S. broadband provider, degraded 25\% from over 2 Mbps in Oct 2013 to 1.5 Mbps in Jan 2014.

To sustain traffic growth, APs need to upgrade network infrastructures and expand capacities; however, their incentives depend on the business model and the corresponding mechanism used to monetize bottleneck bandwidth, which is crucial to the viability of the current Internet model in the future.
A general approach used by APs is to differentiate services and prices, e.g., APs provide premium peering \cite{faratin2007complexity} options for CPs to choose and multiple data plans for end-users with different data usage to choose. However, the former can only be implemented with large CPs via peering agreements, while the latter does not guarantee the performance of end-users in any sense.
The bandwidth allocation is typically a function of the application endpoints, and is traditionally embodied as part of transport layer's congestion control mechanism. TCP CUBIC and BBR are the most popular protocols control the majority of Internet traffic. However, both of them strive for efficient utilization of the bandwidth, while being unaware of the negatively biased user’s Quality of Experience (QoE) affected by Round-Trip Times (RTT) and network buffer size.
Consequently, there exists a fundamental mismatch  between the differentiated services and the underlying resource allocation that differentiates for predictable performance. 

To resolve this mismatch, we consider a class-based differentiation approach, under which CPs and users can choose a {\em service class} (SC) to join. 
We propose \diffperf, a dynamic  performance differentiation framework, at APs vantage point to manage their bottleneck bandwidth resources in a principled and practical manners.
From a macroscopic perspective, \diffperf dynamically allocates bandwidth to each SC according to the changing number of active flows in each SC,
by maximizing the weighted $\alpha$-fair utilities, which enables APs to trade-off fairness.   
Nevertheless, as the users in the same service class might not perceive a fair quality due to the consequences of the complex interaction between the transport protocol and the inherent network conditions such as heterogeneous RTTs and buffer sizes, as shown in our experimental explorations and known by conventional wisdom \cite{bollen1991conventional}.
Thus, at a microscopic level, \diffperf uses a new performance-aware mechanism, called ($\beta,\gamma$)-fairness to further optimize and  
make more fine-grained bandwidth allocation within each SC, so as to more efficiently utilize the aggregate capacity and achieve fairer performance for flows. 
Our main contributions are as follows:
\begin{enumerate}
\item[1. ] 
We derive the closed-form bandwidth allocation solution and show that this solution achieves guaranteed performance differentiation in terms of controllable ratios of the average per-flow throughput across the different SCs. 
\item[2. ] 
Within each SC, we present ($\beta,\gamma$)-fairness and a neat statistical method to differentiate and isolate flows automatically based on their achieved throughput, to mitigate the bias brought by the TCP protocol due to its interaction with network latency (i.e., RTT) and buffer size.
\item[3. ] By leveraging SDN capabilities, we develop a native OpenDaylight (ODL) control plane application that dynamically manages network resources, including tracking flows, inquiring flow statistics and allocating bandwidth capacity. 
Furthermore, to measure the impact of network buffer sizes, we also implement \diffperf on programmable Barefoot Tofino switch which allows flexible buffer sizing and enables fine-grained and flexible line-rate telemetry.
\item[4. ] We carry out comprehensive evaluations of \diffperf from an application perspective for DASH video streaming, as a mainstream accounts for the majority of Internet video traffic.

\end{enumerate}
We believe that \diffperf demonstrates a new avenue for APs to differentiate and optimize the performance of video flows and corresponding perceived user QoE so as to better monetize their bottleneck network resources. This will further incentivize APs to deploy more bandwidth capacity to accommodate the growth of Internet content traffic. 
 \section{The \diffperf Framework}
\label{diffperf}

In this section, we present the \diffperf framework in a top-down manner. We first describe how \diffperf allocates bandwidth capacity among the SCs, based on an optimization approach. We will derive closed-form allocation solution and show its feature of guaranteed performance differentiation. 
We then discuss the performance issues due to the consequences of TCP congestion control mechanism that responds to the heterogeneity of flows' RTTs and network's buffer sizes. To solve this problem, we show how \diffperf classifies flows and optimizes bandwidth allocation within each SC.
\eat{
Currently, the fairness of any form is difficult to be truly implemented by service providers.
The reason is due to the conflict and mismatch between intended business model of the service provider and the transport protocol operations in the behalf of the content providers. 
More specifically, if the ISP wants to allocate certain amount of bandwidth for all users belonging to certain defined subscription plan, that would not be easy to be achieved due to the uncontrolled congestion operations managed by the transport layer. Most of the senders' congestion algorithms rely on RTT to decide the amount of data to send to the receivers. At the bottleneck, when all receivers compete for the bandwidth, short-RTT receivers will presumably allocate a large portion of the buffer and thus grab higher amount of bandwidth than long-RTT ones. In this way, the service provider could not guarantee that users within the same service class could share the aggregate bandwidth fairly. In this paper, we present an optimization framework based on SDN which is able to ensure high precision of predictable performance for users and thus realizing precise business strategy for service provider.
}
\subsection{Inter-Class Bandwidth Allocation} 
We consider an access provider that offers a set $\mathcal{S}$ of service classes over a bottleneck link with capacity  $C$.
We denote the set of active flows in any service class $s\in \mathcal{S}$ by $\mathcal{F}_s$ and the cardinality of $\mathcal{F}_s$, i.e., the number of flows in class $s$, by $n_s$. 
To differentiate the performance for flows in different service classes, the access provider  needs to allocate appropriate amount of bandwidth to each service class. To accomplish this in a principled manner, we formulate the bandwidth allocation as an optimization of the allocation  $\mathbf{X} = (X_s:s\in\mathcal{S})$ that solves a general utility maximization problem as follows. 
\begin{align}
&\underset{\mathbf{X}}{\text{max}}\quad \sum_{s \in \mathcal{S}} n_s U_s\left(\frac{X_s}{n_s}\right) \label{eq:optimization problem 1}\\
&\text{s.t.} \quad\ \ \underset{s \in \mathcal{S}}{\sum }X_s \le C \ \ \text{and} \ \ X_s \ge 0, \ \forall s\in \mathcal{S}. \label{eq:constraint 1}
\end{align}
Under the link capacity constraint~(\ref{eq:constraint 1}), the above mathematical program tries to maximize the aggregate utility over all service classes, where for each service class $s$, it counts the number of flows $n_s$ multiplied by the  per-flow utility $U_s(X_s/n_s)$ over the average capacity $X_s/n_s$ allocated to each flow. 
In particular, we adopt and generalize the well-known weighted $\alpha$ fairness family of utility functions \cite{srikant2012mathematics} as follows.\eat{
\begin{equation}
  U_s(x) = \begin{cases} 
      \vspace{0.05in}
      w_s \log {x} & \text{for} \quad \alpha = 1, \\
      w_s \displaystyle \frac{{x}^{1-\alpha}}{1-\alpha} & \text{for} \quad \alpha \ge 0 \ \text{and} \ \alpha \neq 1.  
   \end{cases}
   \label{eq:Eq 2}
\end{equation}
}
In this family of utility functions, each service class $s$ will be assigned a weight $w_s$ that indicates the relative importance of the service class, resulting in differentiated per-flow bandwidth allocation across the service classes. 
By controlling the parameter $\alpha$, the access provider can express different preferences over various notions of fairness. 
When $\alpha$ approach $0$, the utility tends to be measured purely by the allocated bandwidth; when $\alpha$ approaches $+\infty$, the solution  converges to the weighted max-min fair allocation among the flows. In particular, a trade-off of a weighted proportional fair solution can be obtained by solving the optimization problem when $\alpha$ is set to be $1$. 
Thus, besides the differentiation factor $w_s$ among service classes, the service operator can choose the value of $\alpha$ to tradeoff fairness. 



\begin{theorem}\label{thm:closed form}
If an allocation \(\mathbf{X}\) maximizes the aggregate utility over all service classes, it must satisfy
\begin{equation}
\label{eq:closed form}
X_s = \frac{n_s \sqrt[\alpha]{w_s}}{{\underset{s' \in \mathcal{S}}{\sum }n_{s'} \sqrt[\alpha]{w_{s'}} }} \ C,\quad \forall s\in \mathcal{S}. 
\end{equation}
\end{theorem}

Theorem \ref{thm:closed form} provides the closed-form solution of the utility maximization problem. Based on the optimal allocation solution in Equation (\ref{eq:closed form}), we derive the ratio of the average per-flow capacities of any two service classes \(s,s'\in \mathcal{S}\) as
\begin{equation}\label{eq:ratio}
\left(\frac{X_s}{n_s}\right) : \left(\frac{X_{s'}}{n_{s'}}\right) = \sqrt[\alpha]{\frac{w_s}{w_{s'}}}.
\end{equation}
This result implies that performance differentiation is achieved by enforcing a fixed ratio for the per-flow bandwidth capacity across SCs, which is controlled by the weights~\(w_s,w_{s'}\) and the fairness parameter~\(\alpha\).
Equation (\ref{eq:ratio}) explicitly  shows that the optimal solution 
effectively allocates  a higher average per-flow capacity in the service class that has a larger weight, 
which is desirable and expected for the better service class.
In particular, we also see that when $\alpha$ is set to be $1$, the weighted proportional fair allocation leads an average per-flow allocation that are proportional to the weights of the SCs. 

\subsection{Intra-Class Bandwidth Allocation}
\label{subsec:intra-class}

\eat{
\begin{figure*}
    \centering
    \subfigure[RTT unfairness and impacts of flow isolation]{
    \includegraphics[width=0.30\textwidth]{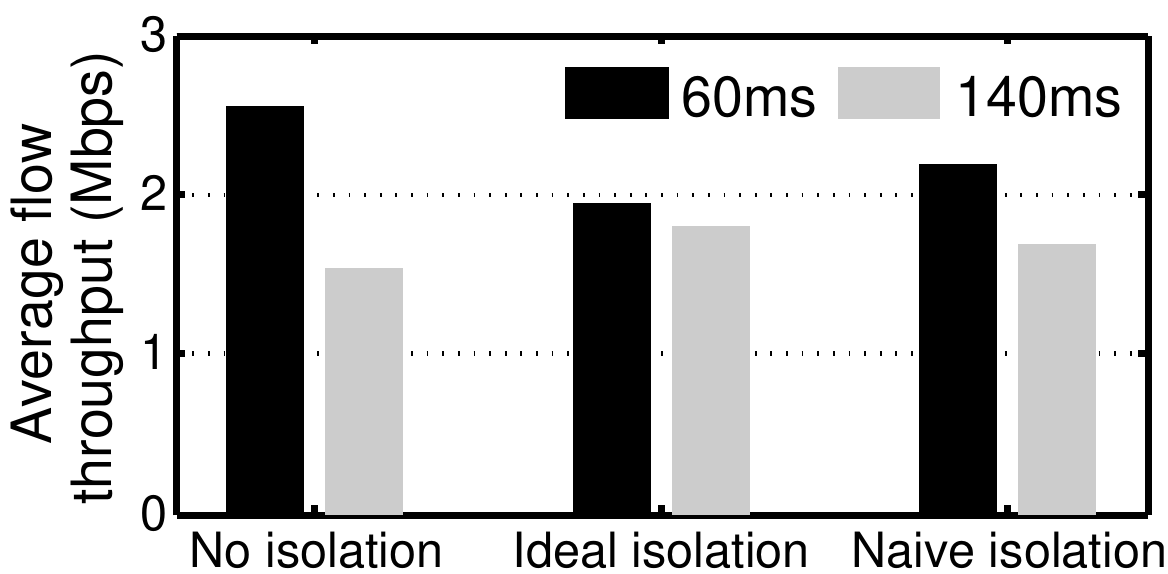}
    \label{fig:intuition}
    }
    \subfigure[Extra delay incurred by SDN control plane]{
    \includegraphics[width=0.30\textwidth]{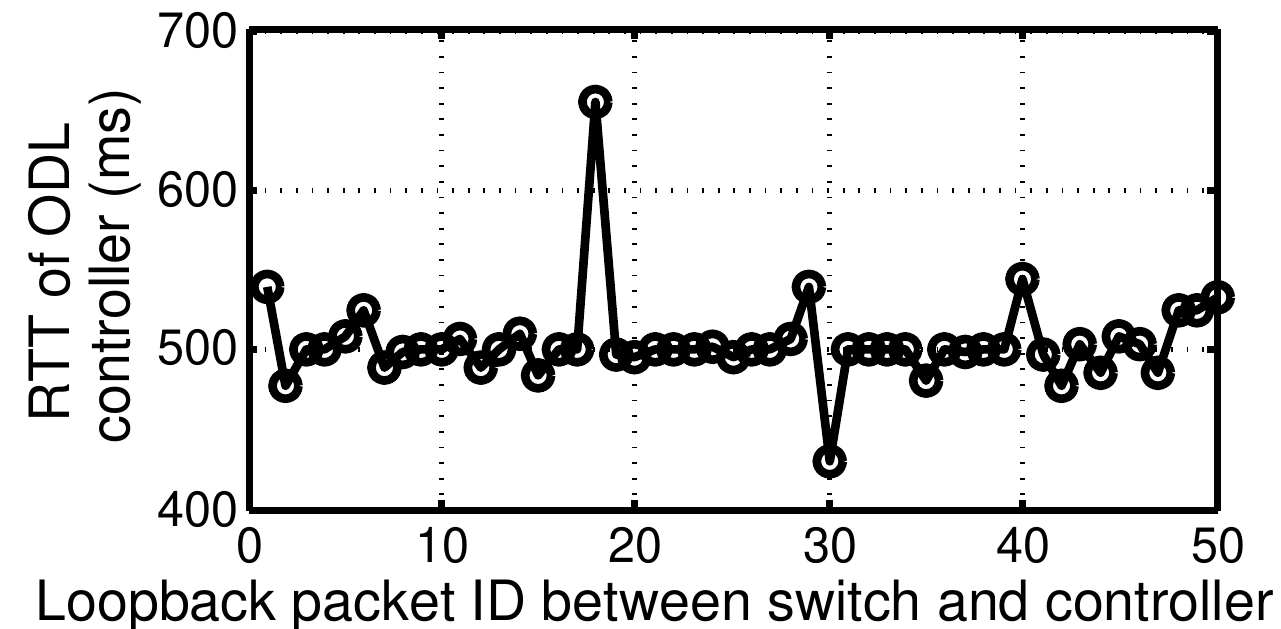}
    \label{fig:ID_RTT}
    }
    \subfigure[Impact of RTT on 
    DASH throughput]{
    \includegraphics[width=0.30\textwidth]{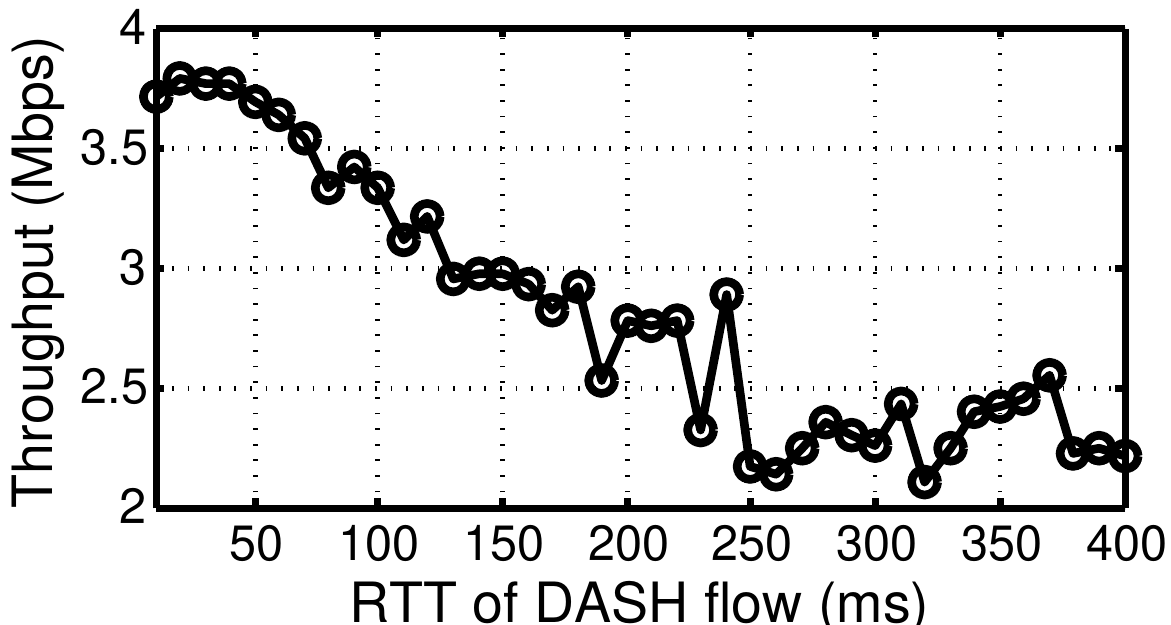}
    \label{fig:RTT_throughput}
    }
    \caption{RTT measurements}
    \label{fi}
\end{figure*}
}
\textbf{Motivation:}
Given $X_s$ amount of bandwidth capacity allocated to the $n_s$ flows in SC $s$, each flow $f\in\mathcal{F}_s$ is expected to achieve an average throughput of $X_s/n_s$. However, the actual throughput achieved, denoted by $x_f$, might be significantly less than the mean.
This can adversely affect the QoE that the corresponding user perceives. 
At the last-mile bottleneck, parameters such as RTT, TCP congestion control algorithm (e.g., CUBIC v/s BBR), and buffer size affect the performance of flows~\cite{hock2017experimental,mathisbuffer,CP-fairness-2019}. The heterogeneity of RTTs experienced by the flows as well as the interaction of the TCP-based congestion control mechanisms that respond to the RTTs and network buffer size differently, lead to multiple  competing flows achieving different throughput.
We analyzed the performance of 100 competing DASH flows on a testbed, where all flows run TCP BBR and share a bottleneck link connecting to a DASH server. The bottleneck link capacity is set to 120Mbps, and 30\% of the flows experience relatively longer RTTs than the rest. We run the experiments by changing one of the key parameters, i.e. the network  buffer size. The experiment results show that the average stalling time of DASH flows at 10MB network buffer size, is 35\% higher than that of DASH flows when the network buffer size is 1MB.
However,  by
``isolating'' flows that perceived dissimilar QoE at the last-mile bottleneck link, we observed that average stalling time of DASH flows is reduced by 50\% and 25\% at the buffer sizes of 1MB and 10MB, respectively, thereby improving the overall QoE significantly.

Motivated by this observation, we propose a practically scalable solution to classifying similar flows into sub-groups and isolating them into separate sub-classes by allocating appropriate amount of bandwidth to them within each SC.
Next, we describe 1) a flexible statistical method that \diffperf uses to classify flows within a SC, and 2) the intra-class bandwidth allocation used by \diffperf for sub-group isolation. 

\subsubsection{Flow Classification and Isolation}
By relying on QoE as a similarity metric to classify the flows, clearly this choice requires an explicit feedback via the receiver to the AP vantage point, which is difficult to afford in practice. We therefore want to leverage SDN functionalities to find other metrics to use at the vantage point. The first metric that comes to mind is RTT. However, the use of real-time RTT samples could not be taken solely as indicators of performance issues without other information such as underlying congestion control mechanism, buffer size~\cite{hock2017experimental}, and packet route security. 
Even if we make assumptions of the availability of these information, measuring flow RTT at the APs is unreliable. Measuring RTT at the SDN control plane inflates a variable and high RTT based on our measurements in ODL control plane, while measuring it in the SDN data plane~\cite{chen2020measuring} may not scale well due to the memory space constraints.
Instead, we emphasize that throughput of TCP flows is the appropriate and robust metric that indicates the collective impact of the interaction of network parameters to the user-perceived performance. 
Next we show how to utilize the throughput measure as a proxy to determining whether flows are similar to each other and to identify effectively which flows are affected.
Because the number of groups and the number of flows for each group may change and are not known in real scenarios, we adopt general statistical metrics for classification. 
Given the achieved throughput $x_f$ of the flows $f\in\mathcal{F}_s$ in any SC $s\in\mathcal{S}$, the mean and standard deviation of the flows' throughput are defined as
$\bar{x}_s = \frac{1}{n_s} \underset{f \in \mathcal{F}_s }\sum x_f \quad \text{and} \quad \sigma_s = \sqrt{\sum_{f \in \mathcal{F}_s } \frac{(x_f - \bar{x}_s )^2}{n_s}}$. 
Because the achieved throughput $x_f$ of each flow depends on  the number $n_s$ of competing flows and their characteristics, 
buffer size, and the allocated capacity $X_s$ that ultimately determines the network congestion imposed on the SC, instead of using absolute throughput thresholds to classify flows, we adopt the following statistical metric that orders and measures the relative throughput values among all flows in the same SC. 

\begin{definition}
Given the mean \(\bar{x}_s\) and standard deviation \(\sigma_s\), the standard score of a flow \(f\)'s throughput is defined by
$z_f = (x_f - \bar{x}_s)/\sigma_s.$
\end{definition}

When a flow's throughput is above (or below) the mean, its standard score or $z$-score is positive (or negative, respectively).
This $z$-score captures the signed fractional number of standard deviations by which it is above the mean value. 

Without loss of generality, we divide a set $\mathcal{F}_s$ of flows into two sub-classes:  lower sub-class $\mathcal{F}_s^L$ and upper sub-class $\mathcal{F}_s^H$, based on each flow's $z$-score compared with a pre-defined threshold $\beta$, where 
$\mathcal{F}_s^L(\beta) = \{f\in\mathcal{F}_s: z_f< \beta\} \quad \text{and} \quad \mathcal{F}_s^H(\beta) =\mathcal{F}_s \backslash \mathcal{F}_s^L(\beta).$
The set $\mathcal{F}_s^L$ contains the flows that achieved the lowest throughput values (i.e., the negatively affected flows). 
Thus, our goal is to identify them
so that we isolate and allocate appropriate amount of bandwidth to them accordingly.
We use a non-positive value of $\beta$ to capture flows whose throughput are $|\beta|$ deviations lower than the achieved average $\bar{x}_f$. 
Because the set  $\mathcal{F}_s^L$ grows monotonically with the parameter $\beta$, i.e., $\mathcal{F}_s^L(\beta_1) \subseteq \mathcal{F}_s^L(\beta_2) \ \forall \beta_1 <\beta_2$, a smaller value of $\beta$ makes a more conservative decision on the 
lowest throughput flows, avoiding mis-classifications.
We will further study how the values of $\beta$ affect the performance of flows in a later section via experimental evaluations.

\subsubsection{Bandwidth Allocation Model}

After classifying flows in each SC into two sub-groups, we isolate them into two sub-classes and determine how much bandwidth $X_s^L$ and $X_s^H$ to allocate for each sub-class. To fully utilize bandwidth capacity, our solution needs to  satisfy $X_s^L + X_s^H = X_s$.

The throughput of some flows 
might be naturally low and might not be able to achieve the targeted throughput $X_s/n_s$ even allocated that amount of capacity.
As a result, enforcing the per-flow allocation of $X_s/n_s$ will result in resource wastage.  
The key question to answer is how much per-flow capacity we should allocate to the flows $\mathcal{F}_s^L$, whose 
innate throughputs are less than what is needed to achieve the average throughput $\bar{x}_s$ or to utilize the per-flow allocated capacity $X_s/n_s$ in theory. Since these flows might not be able to achieve the average throughput, the per-flow allocation should be no higher than $X_s/n_s$.  
On the other hand, by isolating negatively affected flows from high-throughput flows (i.e., the flows $\mathcal{F}_s^H$ that cause the performance issues of flows $\mathcal{F}_s^L$), 
we expect them to achieve higher throughput than what are being achieved; and therefore, we should allocate more capacity for the set $\mathcal{F}_s^L$ of flows than their aggregate achieved throughput. 
To this end, we allocate the average amount of bandwidth capacity for the per flow of set $\mathcal{F}_s^L$ as 
\begin{align}
\frac{X_{s}^L(\beta,\gamma)}{|\mathcal{F}_{s}^L(\beta)|} = \gamma \frac{\sum_{f\in \mathcal{F}_s^-}x_f}{|\mathcal{F}_s^- |} + (1-\gamma)\frac{X_s}{n_s},
\label{eq:Eq 8}
\end{align}
where we define the set of flows whose throughput are below the mean $\bar{x}_s$ by $\mathcal{F}_s^- \triangleq \{f\in\mathcal{F}_s: x_f<\bar{x}_s \}$ and introduce a parameter $\gamma \in [0,1]$ to control the allocated capacity flexibly.
In particular, for one extreme of $\gamma = 1$, the solution allocates the average throughput of the set $\mathcal{F}_s^-$ of flows as the per-flow capacity for the 
lower sub-class $\mathcal{F}_s^L(\beta)$, which must be lower than the average throughput \(\bar{x}_s\) and the average capacity \(X_s/n_s\) of all flows. In this case, the per-flow capacity allocated for the lower sub-class $\mathcal{F}_s^L$ is lower than that allocated for the upper sub-class $\mathcal{F}_s^H$, under which resource wastage is reduced and resource is utilized more efficiently.
For the other extreme of $\gamma = 0$, the solution simply isolates the two sub-classes and equally allocates an average capacity \(X_s/n_s\) as the per-flow capacity for both upper and lower sub-classes, under which per-flow fairness is enforced regardless of how efficiently the resource is utilized.
Thus, by choosing the value of $\gamma$ between $0$ and $1$, we can make a trade-off between resource fairness~and~utilization. However, this depends on the interaction of TCP algorithm with network buffer size. As opposed to the shallow buffer, the deep buffer allows low-throughput TCP flows (especially those negatively affected due to heterogeneity of RTTs) to stabilize their transfer. Thus, in the vantage point of deep buffer, if the low-throughput flows were crowded out by others, then they can perform better if $\gamma = 0$. However, this is not the case in shallow buffer that does not allow negatively affected flows to ramp up quickly and perhaps this even leads to lower utilization.

By Eq.(\ref{eq:Eq 8}), we also have the next theorem showing 1) lower bounds of per-flow capacities re-allocated to the lower and upper sub-classes $\mathcal{F}_s^L$ and $\mathcal{F}_s^H$; and 2) the monotonicity of the average throughput of the flows within $\mathcal{F}_s^L$ and the average per-flow capacity re-allocated to the $\mathcal{F}_s^H$ on the parameter $\beta$. 


\begin{theorem}\label{thm:monotonicity}
Given any fixed parameter $\gamma$, for any service class \(s\in \mathcal{S}\), 
1) the average achieved throughput of the flows within the lower sub-class~\(\mathcal{F}^L_s\) is non-decreasing in \(\beta\) and always no higher than the average per-flow capacity re-allocated to ~\(\mathcal{F}^L_s\); 
2) the average per-flow capacity re-allocated to the upper sub-class~\(\mathcal{F}^H_s\) is non-decreasing in \(\beta\) and always no lower than \(X_s/n_s\).
\end{theorem}

Theorem \ref{thm:monotonicity} states that as the parameter \(\beta\) increases, the average throughput of the flows $\mathcal{F}_s^L$ of the lower sub-class would also increase because more high-throughput flows would be classified into $\mathcal{F}_s^L$. 
It also tells that this achieved average throughput must be no higher than the the per-flow capacity re-allocated to them in $\mathcal{F}_s^L$.
This property guarantees our design objective of allocating more capacity for the flows in the lower sub-class than their aggregate achieved throughput.
Theorem \ref{thm:monotonicity} also states that as \(\beta\) increases, flows within the upper sub-class $\mathcal{F}_s^H$ would be re-allocated more per-flow bandwidth capacity although fewer flows would be classified into the sub-class.
Thus, service operators can choose the value of \(\beta\) to control the scales of the sub-classes and both \(\beta\) and \(\gamma\) to control the bandwidth capacity allocated to the flows of the two~sub-classes, which we refer to it as (\(\beta\),\(\gamma\))-fairness. 

Before we close this section, we would like to emphasize that although \diffperf classifies the flows in each SC into two sub-groups for simplicity, its statistical method of classification and the corresponding bandwidth allocation can be applied in a top-down recursive manner to further split any sub-group for a more fine-grained optimization. 

\section{Implementation}
\subsection{\diffperf Prototype on OpenDaylight with OpenFlow}

We implement \diffperf as an application on the popular industry-grade open-source SDN platform---the OpenDaylight (ODL) controller. 
We particularly develop a native MD-SAL (Model-Driven Service Adaptation Layer) application on ODL which comprises use of different technologies such as OSGI, Karaf, YANG, blueprint container, and messaging patterns as RPC, publish-subscribe, and data store accesses~\cite{ODL}. We skip implementation details for the sake of brevity.
\begin{figure}[htb!]
\includegraphics[width=0.45\textwidth]{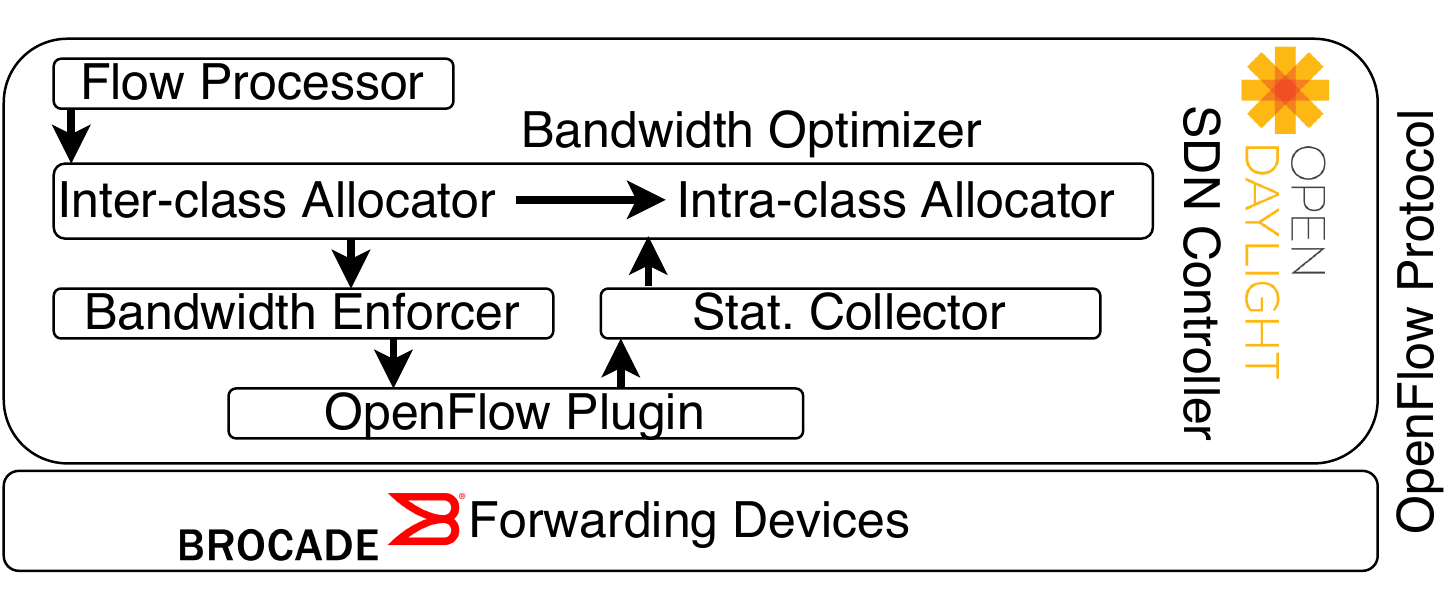}
\centering
\caption{\label{fimp}\diffperf implementation}
\label{fig:implementation}
\end{figure}
Figure~\ref{fig:implementation} depicts the implementation structure of \diffperf on ODL. There are four main modules: {\em Flow Processor}, {\em Bandwidth Optimizer}, {\em Bandwidth Enforcer} and {\em Stats Collector} and are interconnected as shown in the figure; we briefly describe them below.

\eat{
Next, we briefly present components of network control application including flow processor, statistics collector, bandwidth optimizer, and bandwidth enforcer modules. 
}

\subsubsection{Flow Processor}

The {\em Flow Processor} module on the ODL performs two primary functions. First, it assigns user-specified service classes to newly joining flows in the network. We use the YANG modeling language~\cite{yang} to define the service classes. Second, the module carries out regular flow maintenance; i.e., the flow processor inserts new flows into the data store, determines the pre-defined service class and assigns the corresponding weight to new flows, removes inactive or completed flows, etc.

\eat{
For the bandwidth optimizer to determine the optimum bandwidth for the flow, the flow information should be known as a first step, such as service class it belongs to, weight of service class, and desirable optimization allocation. 
For this purpose, we describe a basic structure of a service class based on YANG modeling language that includes above flow information. At runtime, the module is loaded into the controller accompanied with its corresponding datastore and access APIs. A user interface is then rendered to flexibly access to datastore via HTTP-based XML or JSON format. Thus, when the user initiates a network flow connection, the flow can be automatically associated with corresponding pre-defined service class information.
}

\subsubsection{{Statistics Collector}}

\diffperf performs in-network performance optimization.
For \diffperf to work, we need to estimate throughput of each active flow. Let $x_f{(t)}$ denote the average throughput of a flow~$f$ until time $t$. By measuring the instantaneous throughput~$\tilde{x}_f$ of flow~$f$ during the past immediate time period $\Delta_t$, the average throughput for the next time period $\Delta_t$ is updated as follows:
$x_f{(t+\Delta_t)} \leftarrow \delta \times x_f{(t)} + (1-\delta) \times \tilde{x}_f.$ 
\label{eq:througput-est}
where $\delta \in[0,1]$, is a weight. 
Having measured the average throughput of the active flows, we use these estimates to group flows into sub-classes so that flows with similar achieved throughput fall into the same sub-class. 

To obtain real-time estimates of the throughput of each active flow as well as link bandwidth, we implement a {\em Stat. Collector} module on ODL. This module registers per-flow rules to pull out measurement information using event-based handlers from the {\em operational data store} in ODL. The data store in turn uses the OpenFlow plugin (as indicated in Figure~\ref{fig:implementation}) to request the switches to report flow measurements; 
The per-flow measures of interest are packet counts, byte counts and duration. 

\subsubsection{{Bandwidth Optimizer}} 

The core part of \diffperf is the {\em Bandwidth Optmitizer} module, which is responsible for inter- and intra-class bandwidth optimization described in Section~\ref{diffperf}. The optimizer runs every $\Delta_t$ interval, getting input from two modules described above - {\em Flow Processor} and {\em Stat. Collector} (see Figure~\ref{fig:implementation}). While the former provides mapping of flows to user-specified service classes, the latter provides real-time measurements on the active flows in the switch.  Given the input information, the inter- and intra-class optimizers are executed; the output of optimization are: (i)~the portion of bandwidth allocated to each service class (SC), and (ii)~the portion of bandwidth for each sub-class within every service class.



\subsubsection{{Bandwidth Enforcer}}

To materialize bandwidth allocation, each sub-class should use its designated bandwidth in an isolated manner. A naive approach to implementing this is to leverage multi-queues at the switch egress port so that each sub-class maps into an isolated queue. However, there are two practical challenges.
First, in commodity switches the number of queues at egress port is usually limited to a small number~\cite{cisco}\cite{brocade}, meaning that the number of available queues could be less than the number flow sub-classes. Second, 
current OpenFlow switches do not expose APIs to update the weight of the queues dynamically. Without this capability (of dynamically changing queue weights), the bandwidth allocated to queues cannot be changed as and when required. 

To overcome both limitations, we leverage the {\em metering} feature available in OpenFlow switches. Instead of defining queues and updating their bandwidth at egress port, we develop a {\em Bandwidth Enforcer} module that essentially does enforcement at the ingress side of the switch. That is, multiple meters corresponding to the number of sub-classes are defined; and based on the output of the {\em Bandwidth Optimizer}, flow rate of each sub-class is attached to a specific meter dynamically. 
The {\em Bandwidth Enforcer} uses the OpenFlow plugins to encapsulate the allocated bandwidth into OpenFlow messages and install them on to the switch(es).

\subsection{\diffperf Prototype with Programmable Data Plane}
Next, we implement another prototype of \diffperf on a lightweight \texttt{C} controller that is connected to Barefoot Tofino programmable switch~\cite{Tofino} which enables flexible buffer sizing and fine-grained line-rate telemetry. We particularly implement statistics building block in the data plane to track number of bytes transmitted by the active flows. Additionally, we re-implement \textit{Bandwidth Enforncer} and \textit{Statistics Collector} modules in the control plane. We leverage the Tofino switch exposed APIs to update the weight of the queues dynamically and configure their sizes by the control plane. The remaining modules are kept the same with minor modifications.

\section{Experimental Evaluation}

We evaluate \diffperf by carrying out experiments on a realistic testbed. We describe the details below. 

\subsection{Testbed setup}

\textbf{OpenFlow Brocade switch experiments:} We set up a testbed for video streaming between a DASH client (i.e., video player) called \texttt{dash.js} and DASH server over an SDN network; 
Our testbed consists of 12 servers, 10 of which are used to host DASH clients and one each for hosting DASH server and ODL controller. The 10 servers running DASH clients are connected to the DASH server such that they compete (for video segments) at a downstream bottleneck link from a SDN-enabled Brocade ICX-6610 24-port physical switch. We evaluated \diffperf in three different scenarios. In Scenario~1 (Section~\ref{s1}) and Scenario~2 (Section~\ref{s2}), each physical server hosts up to $4$ DASH clients, each client runs in a VM, and all clients are connected to the DASH server over a 50~Mbps downstream bottleneck link. For Scenario~3 (Section~\ref{s3}), we scale up the number of DASH clients---each physical server hosts $15$ DASH clients, all run as docker containers and connected to the DASH server over a 200~Mbps downstream bottleneck link.

\textbf{Barefoot Tofino switch experiments:}
The experiments with Tofino programmable switch concentrate on
evaluating the impact of buffer size on the performance of bottlenecked flows and how \diffperf enables switch buffer to perform better (i.e., improve the overall flow performance). 
The evaluations  are carried out with multiple switch buffer sizes: $100$KB, $1$MB, and $10$MB.
Tofino exposes a set of APIs for Traffic Manager applications to manage buffer allocation from both ingress and egress ends. We use \texttt{bf\_tm\_q\_app\_pool\_usage\_set} API to aid in setting buffer size for the queues of the egress port attached to the bottlenecked link. Buffer size is specified in terms of cells, where each cell size is 80 bytes. The buffer precedes a $120$ Mbps bottleneck link that transfers video segments from DASH server to $100$ DASH clients. The results are presented under the last part of Scenario~2 (Section~\ref{s2}).

Except for Scenario~1 (Section~\ref{s1}), assuming majority of flows in the Internet have short RTTs~\cite{farahbakhsh2015far}, we partition the clients into two sets in 70:30 ratio based on the $\text{RTT}_\text{min}$ values configured: the mean and standard deviation of the bigger set are 64ms and 16ms, respectively, and that of the other are 224ms and 32ms, respectively. We use the network emulator {\em netem}~\cite{netem} at the server machines running the DASH clients to set the latency.
For streaming, we use the Big Buck Bunny video sample that lasts for 600 seconds and has been encoded into 3 bitrate levels---1.2 Mbps, 2.2 Mbps and 4.1 Mbps---of equal segments (i.e., each segment is 2 seconds long). Thus, a DASH client can choose the bitrate levels and segments for streaming video, based on the measured congestion level of the network.
We compare \texttt{DiffPerf} against two most popular TCP congestion control algorithms on the Internet~\cite{mishra2019great}; TCP CUBIC~\cite{ha2008cubic} and TCP BBR~\cite{BBR-Jacobson-2016}.
\subsection{Metrics for evaluation}

To evaluate the performance of \diffperf and TCP variants, we use two metrics. One is per-flow average throughput, which in our case corresponds to the average throughputs of all DASH client. 
The other metric of importance is the user-perceived quality-of-experience (QoE). The QoE metric is adopted based on the widely used model proposed by~\cite{yin2015control}, and is expressed as:
$\text{QoE}=\sum_{n = 1}^{N} q(R_n)- \lambda \sum_{n=1}^{N-1}|q(R_{n+1})-q(R_n)|-\mu T_{stall}-\mu_s T_s.$
This QoE definition uses various performance factors such as the average playback bitrate $R_n$ over the total $N$ segments of the video, the average variability of the consecutive segments bitrate represented by the second summation, the duration of rebuffering $T_{stall}$ (i.e., the duration of time the player’s playout buffer has no content to render), and startup delay ($T_s$) (i.e., the lag between the user clicking and the time to begin rendering). As in~\cite{yadav}\cite{yin2015control}, $q$ maps a bitrate to a quality value; $\lambda$ is usually set to one, $\mu$ and $\mu_s$ are set to the maximum bitrate of the video sample. 
We measure QoE for the entire duration of the video.

\begin{figure}
    \centering
    \subfigure[Throughput]{
    \includegraphics[width=0.22\textwidth]{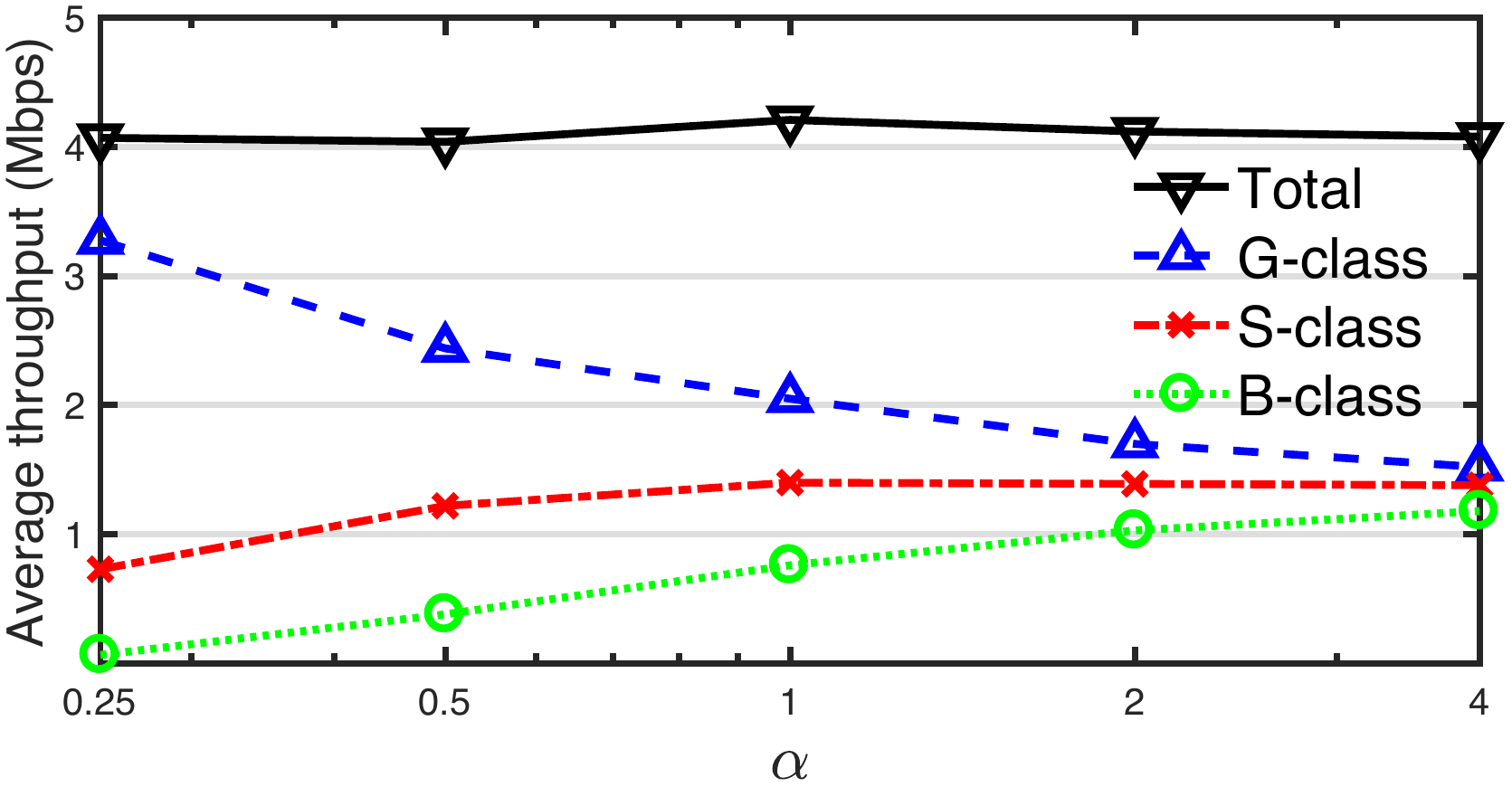}
    \label{alpha_48}
    }
    \subfigure[QoE]{
    \includegraphics[width=0.22\textwidth]{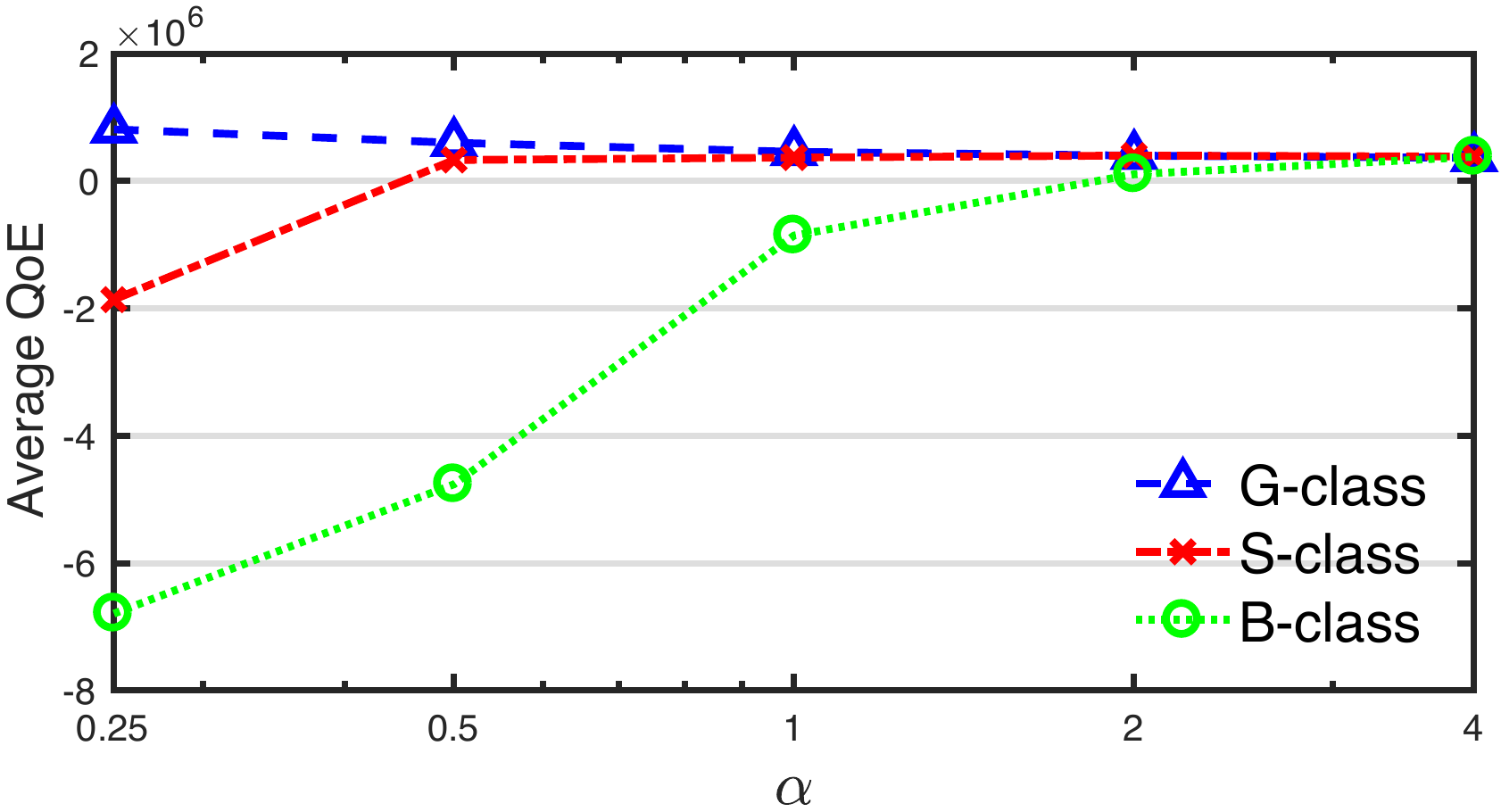}
    \label{alpha_qoe}
    }
    \caption{The Achieved throughput and corresponding QoE of multiple service classes for different values of \(\alpha\)}
    \label{fig:Alpha_throughput}
\end{figure}
\subsection{Results}

\subsubsection{\underline{Scenario 1: Evaluation of inter-class performance}}
\label{s1}
In this scenario we evaluate \diffperf's inter-class bandwidth allocation model. We assume the access provider offers three classes of services: Golden (G), Silver (S) and Bronze (B), with weights of 3, 2 and 1, respectively. We run a set of experiments to evaluate the bandwidth allocated to users of different classes under different values of $\alpha$. We assign 13 DASH clients to each service class, thereby having a total of 39 DASH clients in this scenario. All flows experience homogeneous RTT in this scenario.

Figures~\ref{alpha_48} 
plots, for each SC, the average throughput achieved by all flows in that class, for different values of $\alpha$. Evidently,
the ratios of the estimated average throughput of flows across the service classes closely follow the ratios obtained from our model (refer Eq.~\ref{eq:ratio}). In addition, the average throughput is converging with increasing $\alpha$. 

Figure~\ref{alpha_qoe} plots the average QoE of all flows in each SC.
Observe that the QoE of service class~B is low, when the average throughput achieved (given in Figure~\ref{fig:Alpha_throughput}(a) achieved is low. The QoE of the three service classes converge with increasing value of $\alpha$. With increasing $\alpha$, as resources would be fairly shared among the competing flows, it is expected that higher level QoE also reflects this fair sharing given that the flows have homogeneous RTTs. 

\subsubsection{\underline{Scenario 2: Evaluation of intra-class performance}}
\label{s2}
In this part, we evaluate our proposed performance-aware fairness, ($\beta$,$\gamma$)-fairness. That is, \texttt{DiffPerf}'$\beta$ capability to mitigate the bias brought against the affected flows by the interaction between TCP CUBIC, TCP BBR, flows with heterogeneous RTTs, and switch buffer size. 
Also, we present the flexibility of $\gamma$ in enabling a feature of practical interest, the trade-off between network efficiency and user QoE fairness. Recall, \diffperf uses statistical flow classification and bandwidth allocation to the classified sub-classes to appropriately allocate a higher capacity to the negatively affected flows.

\textbf{~\diffperf based on CUBIC.}
\begin{figure}[hbt!] 
    \centering
    \includegraphics[width=0.30\textwidth]{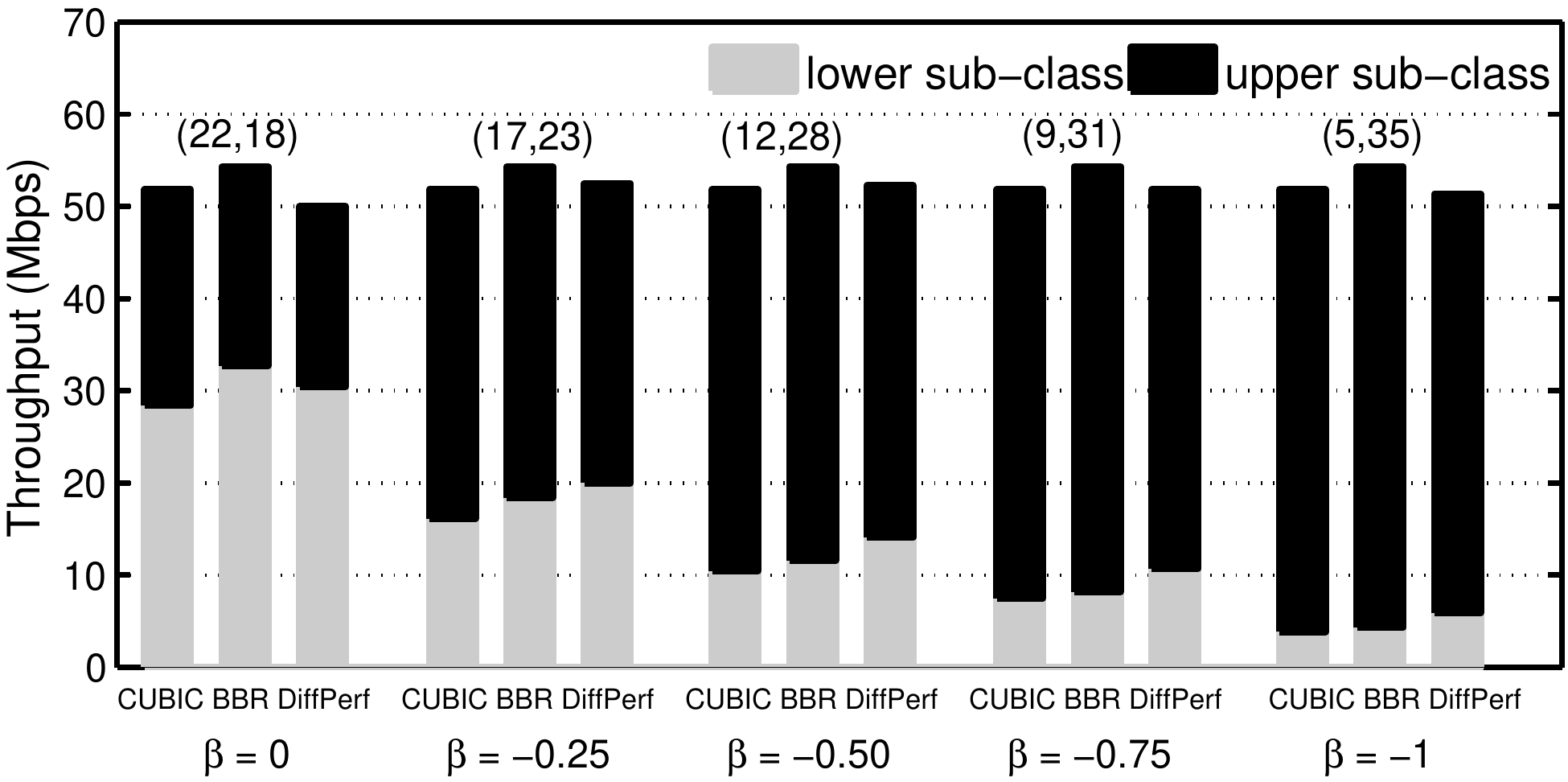}
    \caption{Aggregate throughput of the SC sub-classes flows 
    }
    \label{fig:comp_Beta_throughput}
\end{figure}
\begin{figure}[hbt!] 
    \centering
    \subfigure[Upper sub-class]{
    \includegraphics[width=0.22\textwidth]{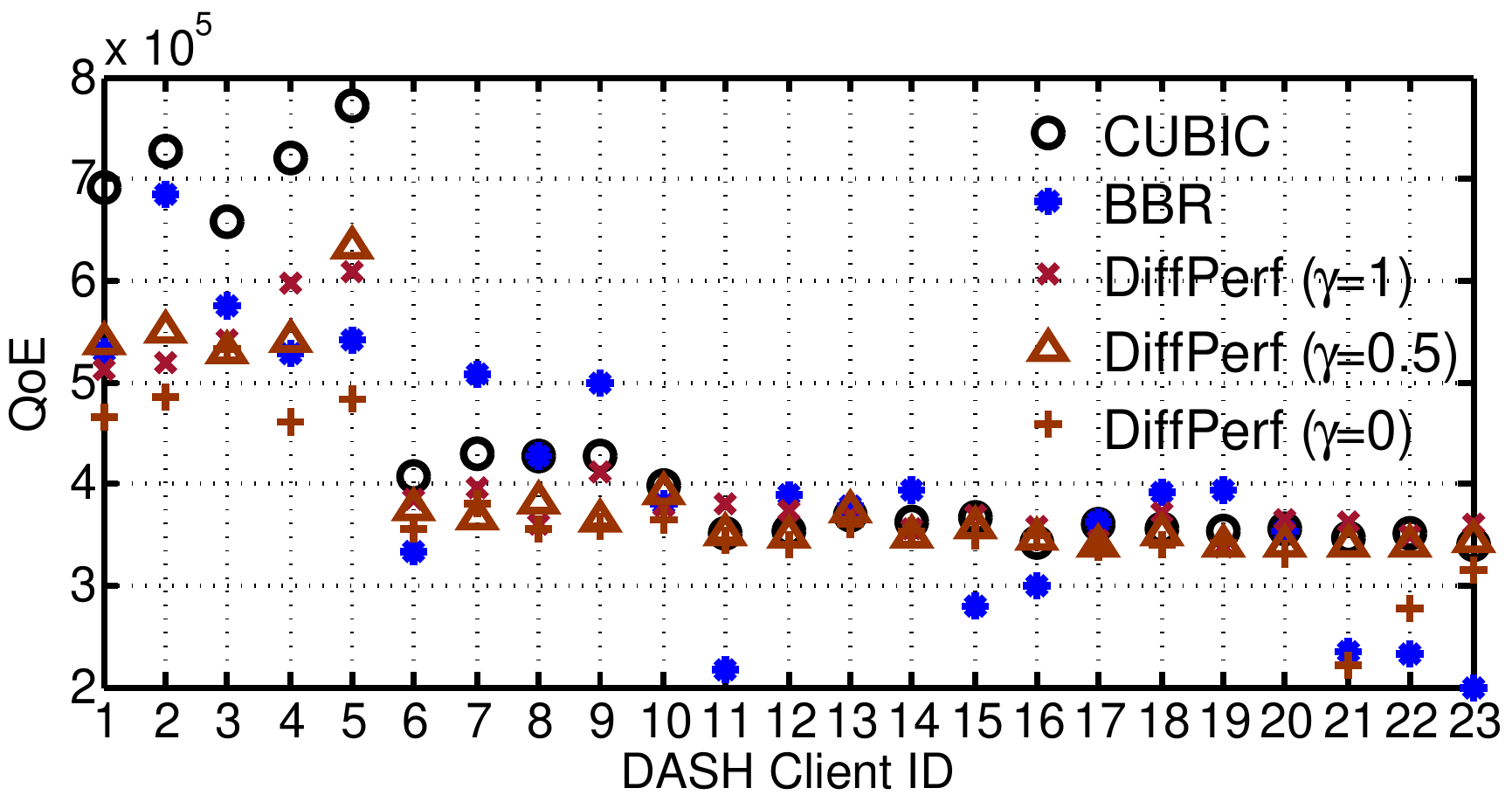}
    }
    \subfigure[Lower sub-class]{
    \includegraphics[width=0.22\textwidth]{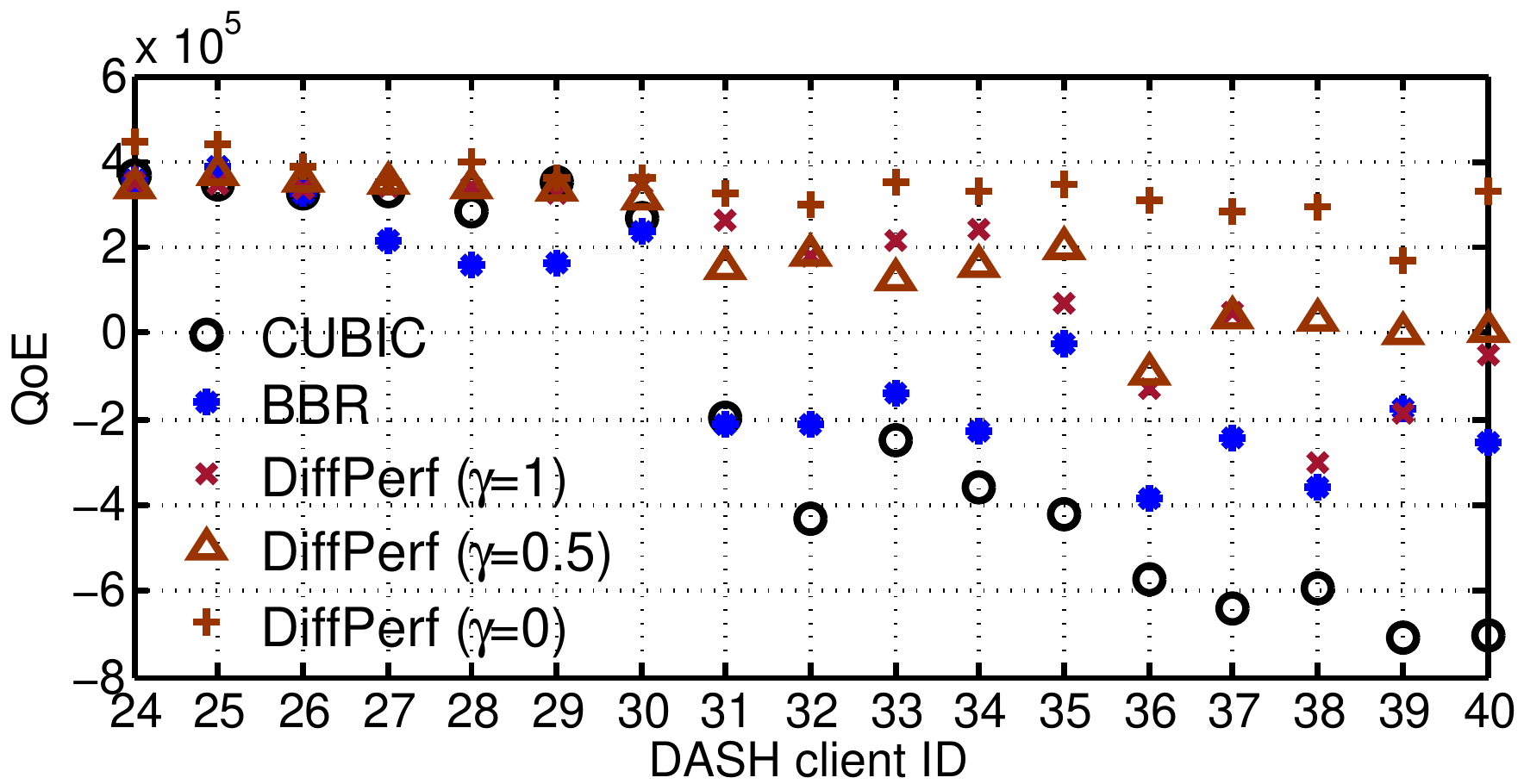}
    }
    \label{fig:gamma_overall_QoE}
    \caption{QoE of the SC sub-classes flows}
    \label{fig:gamma_QoE}
\end{figure} 
In Figure~\ref{fig:comp_Beta_throughput}, we present the aggregate throughput achieved by DASH clients belonging to the two different sub-classes, under TCP CUBIC, TCP BBR and \diffperf over CUBIC. We run experiments for different values of $\beta$. For better clarity, we show the results only for $\gamma = 0.5$. 
The number of flows classified into the lower sub-class based on our statistical classification, for $\beta = 0, -0.25, -0.5, -0.75,$ and $-1.0,$ are 22, 17, 12, 9, and 5, respectively. The figure shows that \diffperf's \textit{isolation} enables the DASH clients in the lower sub-class to achieve higher throughput than both TCP variants, while also being to achieve comparable aggregate throughput as TCP\footnote{Notice that all CUBIC, BBR, and \diffperf achieve aggregate throughput slightly higher than 50Mbps; this is due to the burstiness tolerated by OpenFlow meters. Also, BBR achieves the highest aggregate as it fundamentally does not passively react to packet loss or delay as signals of congestion.}. And $\beta$ gives an access provider flexibility to decide on the flow classification based on a simple and intuitive metric ($z$-score). 
While small number of flows get classified in the lower sub-class, we note that these were also the worst affected onces.
Figures~\ref{fig:gamma_QoE}(a) and~(b) plot the QoE achieved by each DASH client, for $\beta = -0.25$; here we plot for different values of $\gamma$ as well. The former figure plots QoE of the upper sub-class of flows, the the latter plots the same for the lower sub-class of flows. The key observation is that, the flows in the lower sub-class perceive higher QoE under \diffperf than under both TCP variants, and only at the cost of a small number of flows in the upper sub-class. Another observation is that, with lower $\gamma$, \diffperf gives fairer QoE to the clients; \diffperf with $\gamma = 0$ is the most fair. 
\diffperf not only improves the fairness, we calculated the overall QoE values for all flows; and it shows that \diffperf, via flow isolation as well as performance-aware bandwidth allocation, improves significantly the overall QoE compared to TCP solely. It performs $1.86$, $1.62$, and $1.58$ times higher QoE than CUBIC, at $\gamma = 0$, $0.5$, and $1$, respectively; Similarly, it performs $1.63$, $1.42$, $1.38$ times higher QoE than BBR.

\begin{figure}[t]
    \centering
    \includegraphics[width=0.30\textwidth]{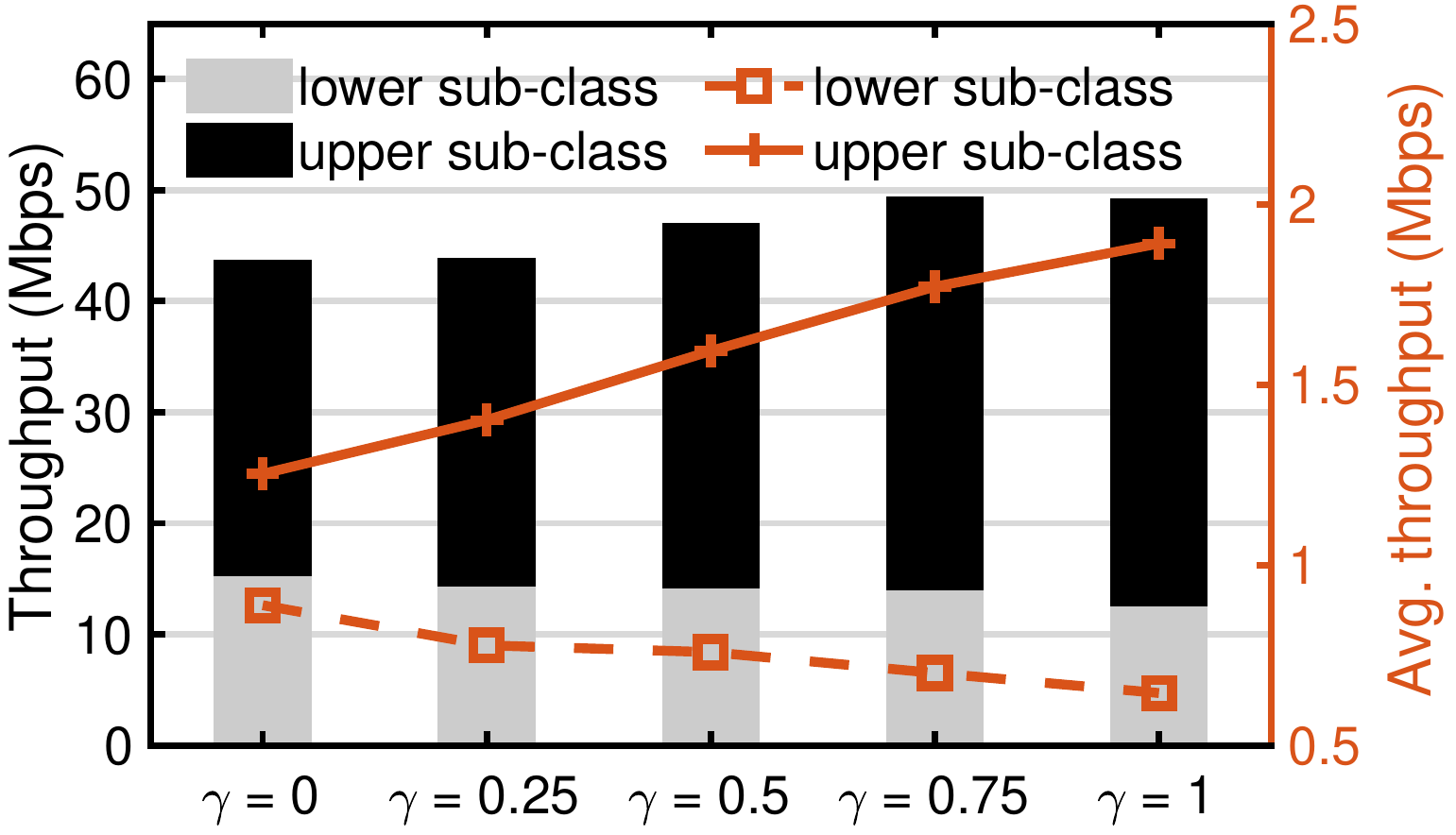}
    \caption{\diffperf fairness-efficiency tradeoff}
    \label{fig:gamma-tradeoff}
\end{figure}

\textbf{Fairness-efficiency tradeoff:} We run experiments for different values of $\gamma$, to analyze the trade-off between efficiency (i.e., bandwidth utilization) and fairness (user-perceived quality fairness), $\beta$ is set to $-0.25$. 
Figure~\ref{fig:gamma-tradeoff} depicts that the aggregate throughput  increases as $\gamma$ increases. Meanwhile, the average throughput of the lower sub-class decreases and that of the upper-class increases, as $\gamma$ is increased from~0 to~1 (refer the second Y-axis). 
Evidently, the parameter $\gamma$ affects the average flow throughput of both sub-classes.
This behavior is due to the fact that, when $\gamma$ approaches 0, the clients are allocated equal bandwidth
regardless of the sub-class; hence the DASH clients tend to achieve higher throughput (subject to their characteristics), and thus the fairness is also improved.
Conversely, when $\gamma$ approaches~1, \diffperf helps the network achieves better bandwidth utilization. \diffperf allocates higher bandwidth to upper sub-class that likely has flows with greater tendency to exploit provisioned network bandwidth, and hence result in better network utilization. 

\begin{figure}[hbt!] 
    \centering
    \includegraphics[width=0.30\textwidth]{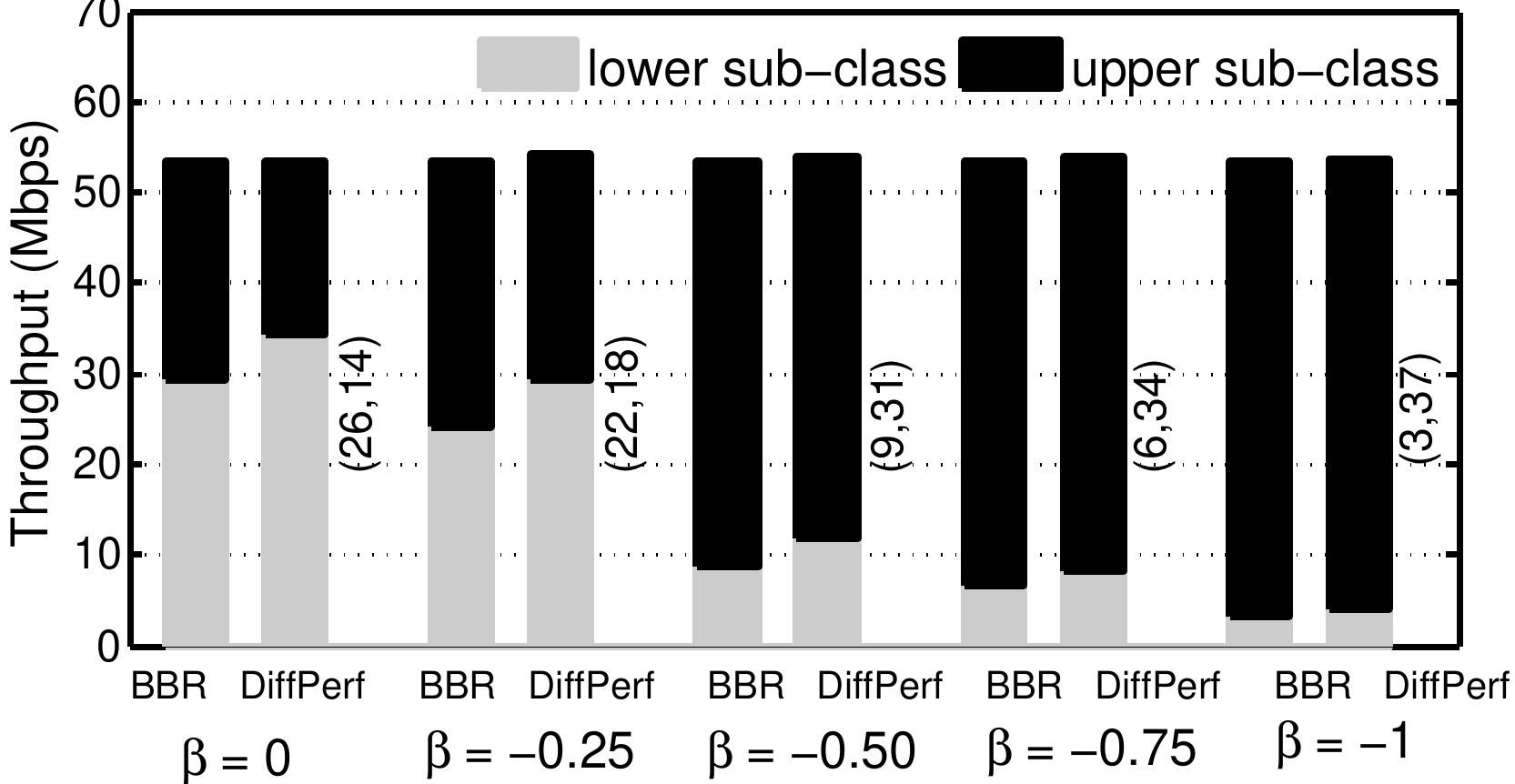}
    \caption{Aggregate throughput of the SC sub-classes flows}
    \label{fig:bbr_thr}
\end{figure}

\begin{figure}[hbt!] 
    \centering    
    \subfigure[Upper sub-class]{
    \includegraphics[width=0.22\textwidth]{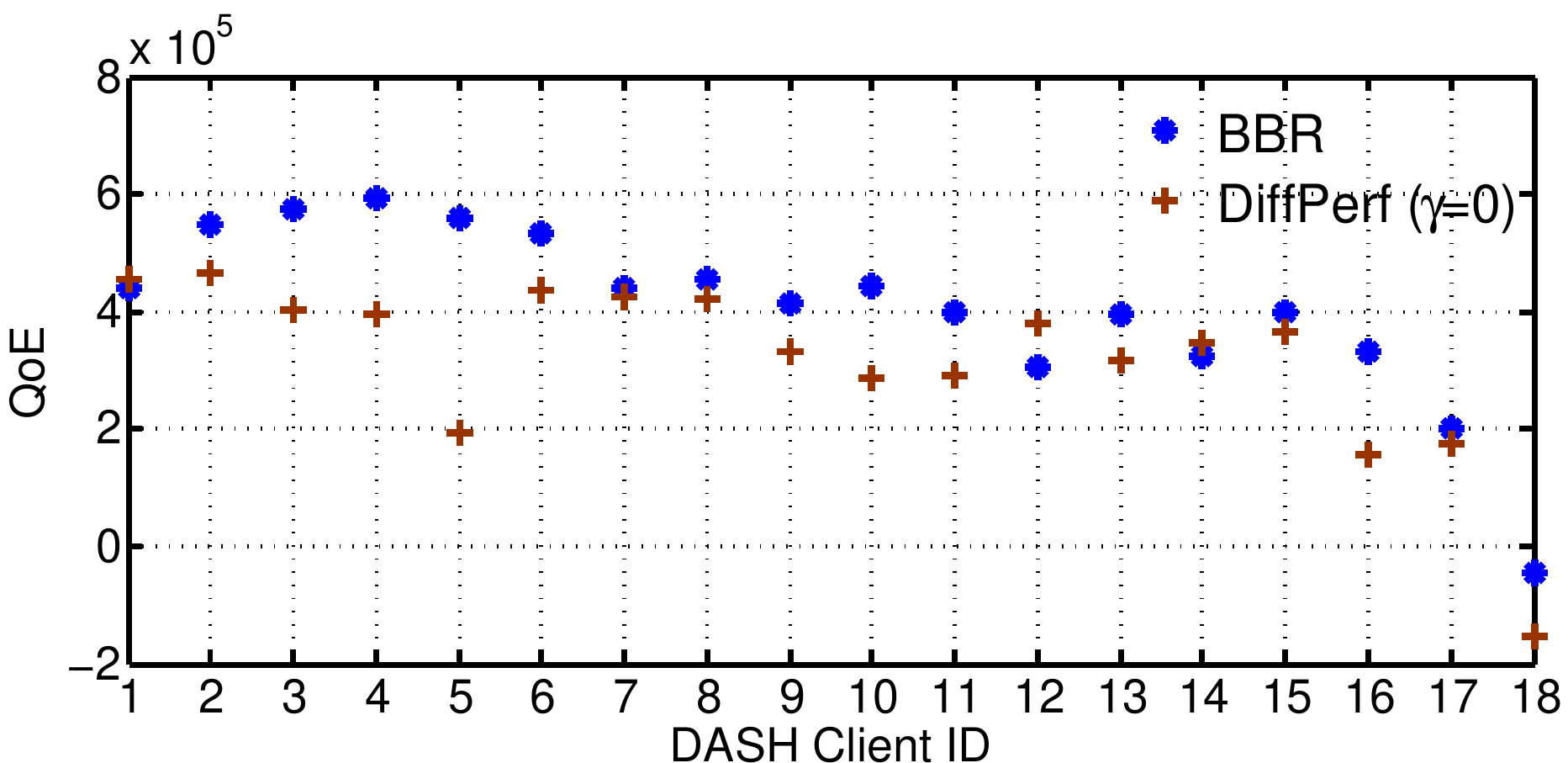}
    }
    \subfigure[Lower sub-class]{
    \includegraphics[width=0.22\textwidth]{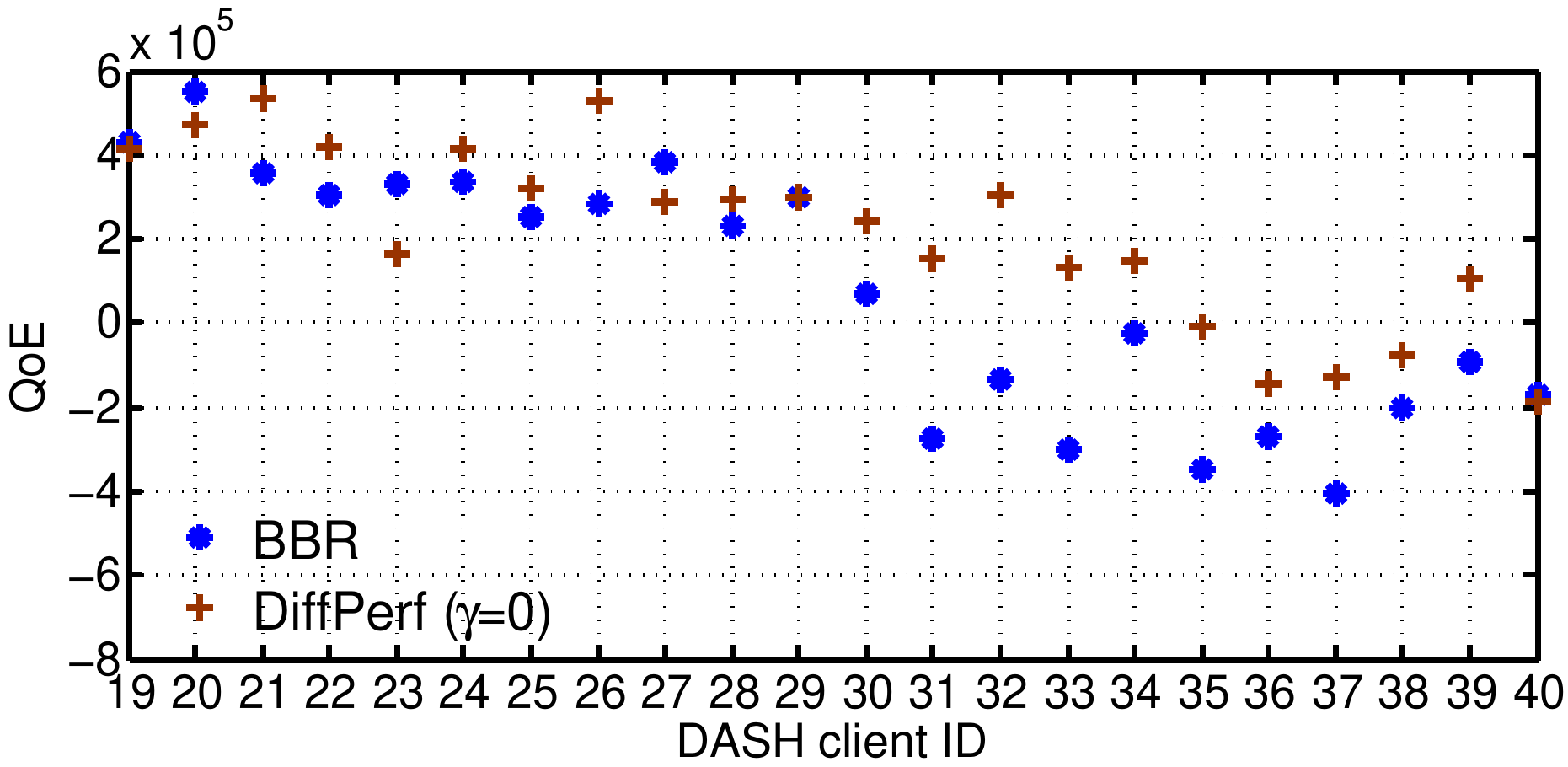}
    }
    \caption{
    QoE of the SC sub-classes flows
    }
    \label{fig:bbr_gamma_QoE}
\end{figure}

\textbf{~\diffperf based on BBR.}
As TCP BBR has recently gained wide-spread attention, \diffperf is evaluated over TCP BBR. 
Figure~\ref{fig:bbr_thr} shows that \diffperf's \textit{isolation} enables DASH clients in the lower sub-class (i.e., the affected flows) to achieve higher throughput than BBR, while also being able to achieve comparable aggregate throughput as BBR. The number of flows classified into the sub-classes are illustrated next to \diffperf bar. For example, at $\beta$ = $0$, $26$ flows are classified into lower sub-class, and $14$ flows to upper sub-class.
Figure~\ref{fig:bbr_gamma_QoE}(a) and (b) plot the perceived QoE of DASH clients in the aforementioned sub-classes. The lower-subclass flows with \diffperf perceive better QoE than with BBR.
    
\begin{figure}[hbt!]
    \centering
    \includegraphics[width=0.45\textwidth]{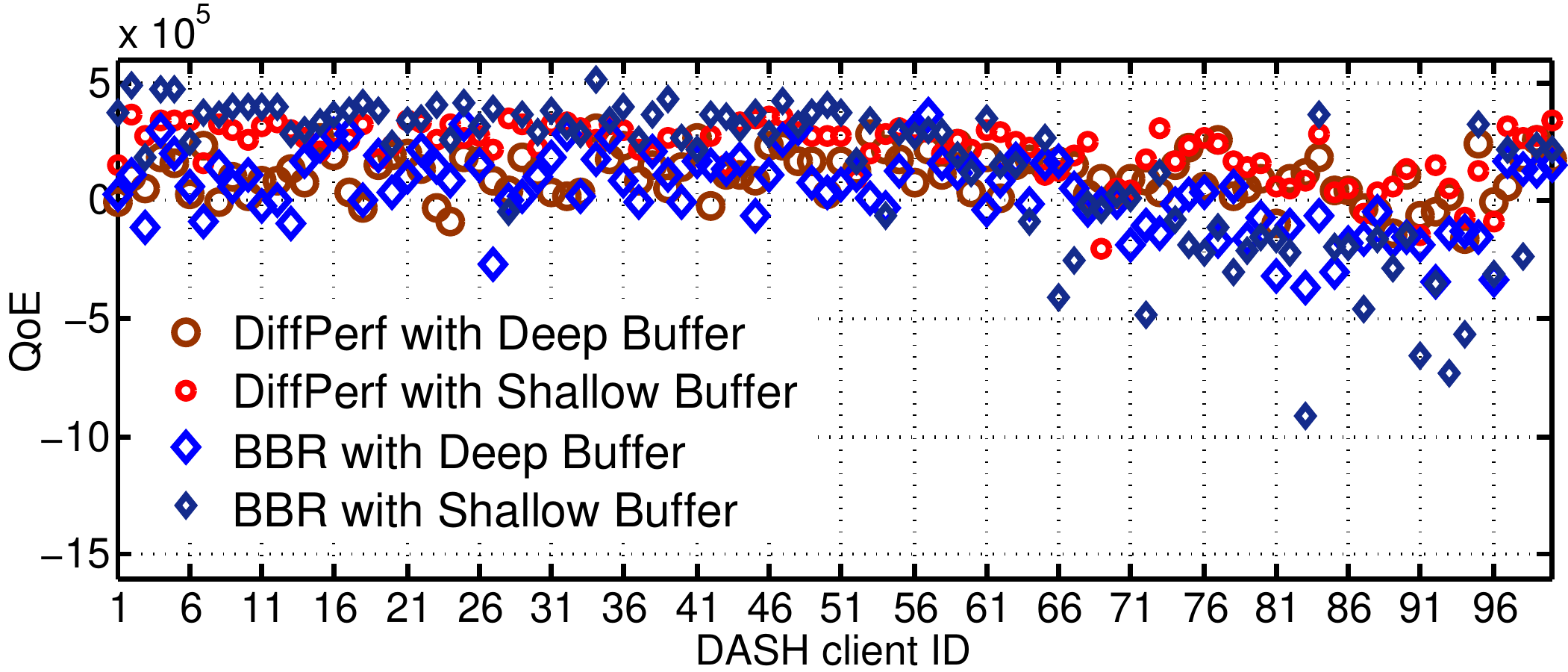}
    \caption{Impact of switch buffer size}
    \label{fig:bufSize}
\end{figure}

\textbf{Impact of buffer size:} Based on the experiments carried out on the Tofino switch, this part presents the impact of buffer size on the performance of bottlenecked flows and the flow optimization achieved using \diffperf.
Note, Tofino switch updates \diffperf with flow statistics every 1s (i.e., the sampling rate). At every interval of $\Delta_t$~=~5s, \diffperf uses the last measured statistics for regulating network flows in the next immediate time interval.
Figure~\ref{fig:bufSize} plots the QoE achieved by $100$ DASH clients competing for $120$Mbps bottleneck link preceded at one experiment by $1$MB shallow buffer  and in the other experiment by $10$MB deep buffer. With shallow buffer, BBR achieves three times higher QoE than with deep buffer. 
We then run same experiments for \diffperf, with $\beta = -0.25$ and $\gamma = 0$; and it achieves $1.58$ times higher QoE than BBR with shallow buffer and $2.6$ times higher QoE than BBR with deep buffer. 
We note that shallow buffer leads to overall better user-perceived quality. However, quality worsens with much smaller buffer size (e.g. $100$KB). The deep buffer might help low-throughput flows, especially those affected by the interaction of TCP with flow RTT, to achieve better QoE
but it increases packet queuing delay. The very shallow buffer (e.g. $100$KB) reduces packet queuing delay but increases packet losses. Hence, both these extreme buffer sizes increase DASH client's average stalling time (i.e., the duration of time the player’s playout buffer has no content to render). With BBR, the client average stalling time is $92.2$, $56.4$, and $76.5$ seconds, while under \diffperf, the client average stalling time is 
reduced to $38.8$, $28.8$, and $58.4$ seconds over buffer sizes $100$KB, $1$MB, and $10$MB, respectively. \diffperf thus demonstrates to be effective in improving the user-perceived quality for multiple buffer sizes. 
Lastly, it is worth noting that from this set of experiments, the buffer size of $1$MB makes better trade-off between queuing delays and packet losses.
 \begin{figure*}[hbt!]  
    \centering
    \subfigure[Arrival-departure of the Clients]{
    \includegraphics[width=0.23\textwidth]{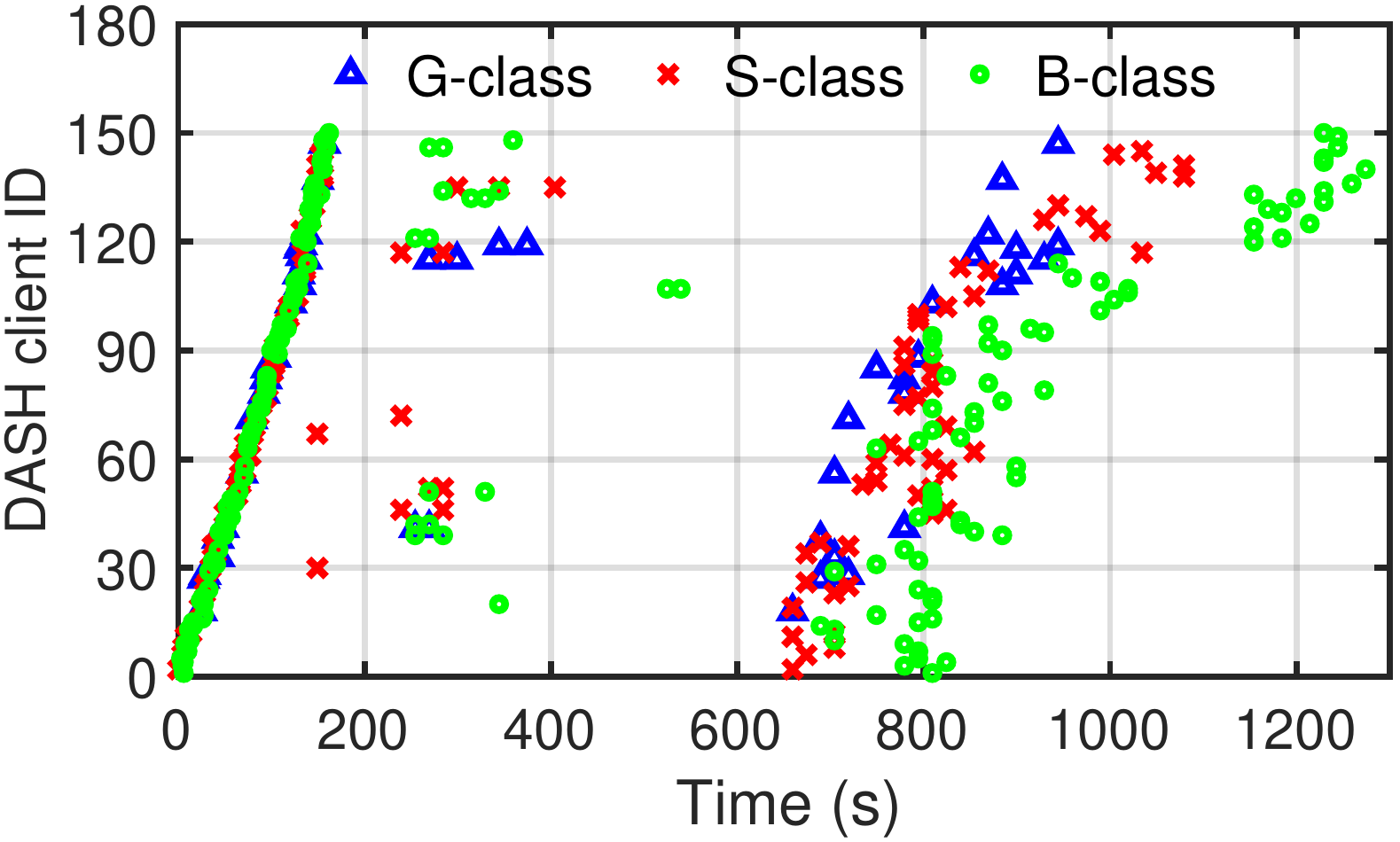}
    \label{fig:arr-dep}}
    \subfigure[G-class active flows]{ 
    \includegraphics[width=0.23\textwidth]{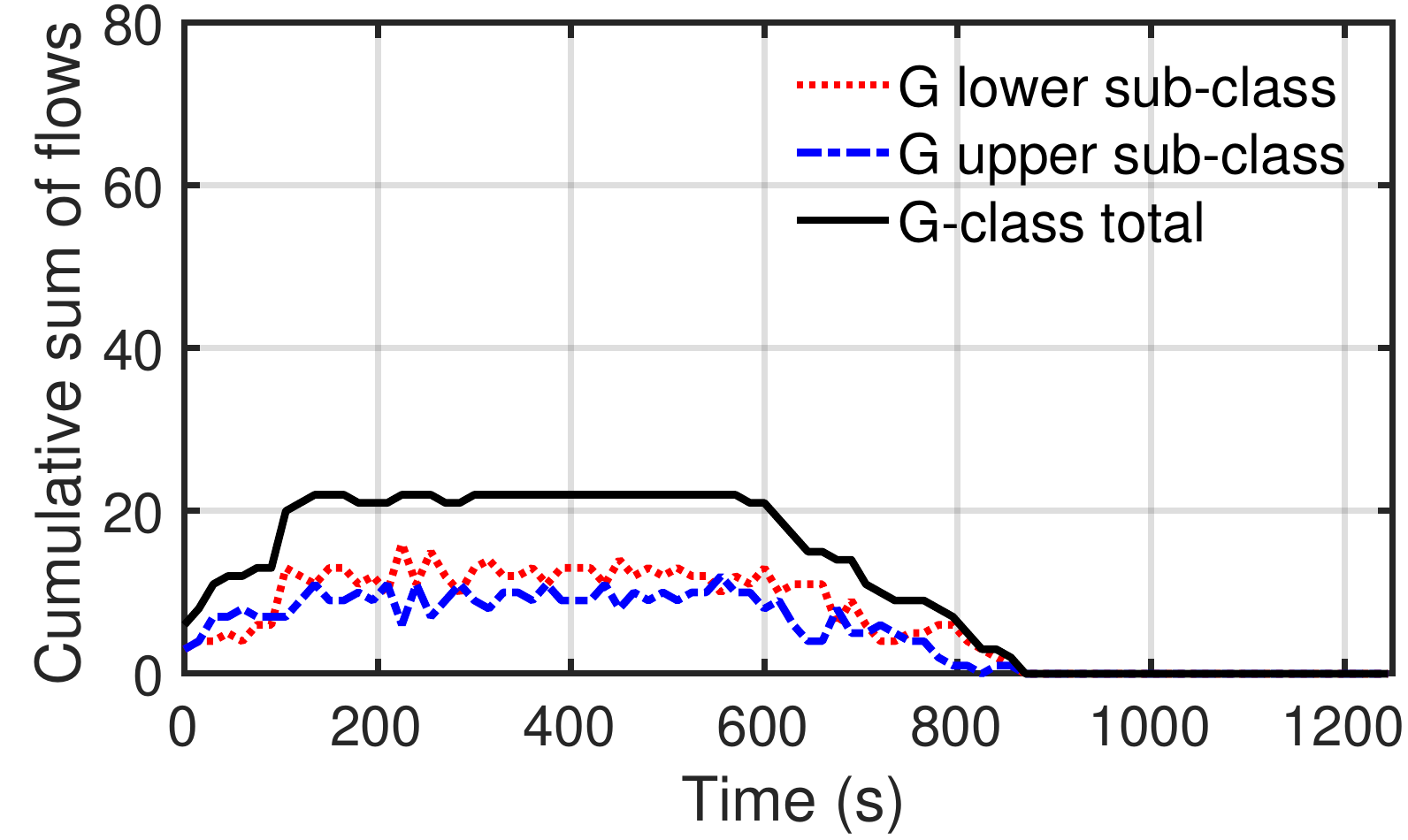}
    }
    \subfigure[S-class active flows ]{
    \includegraphics[width=0.23\textwidth]{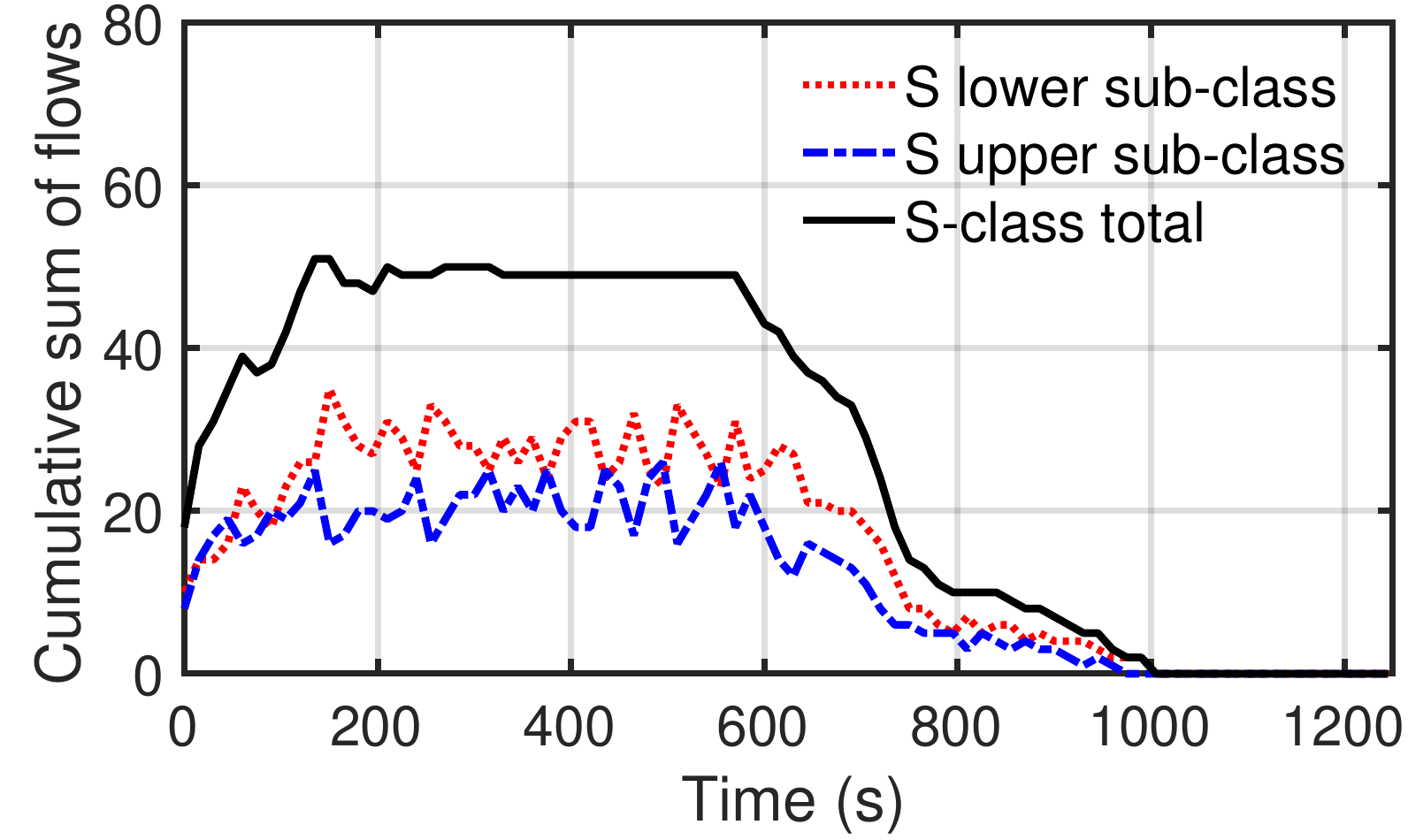}
    }
    \subfigure[B-class active flows]{
    \includegraphics[width=0.23\textwidth]{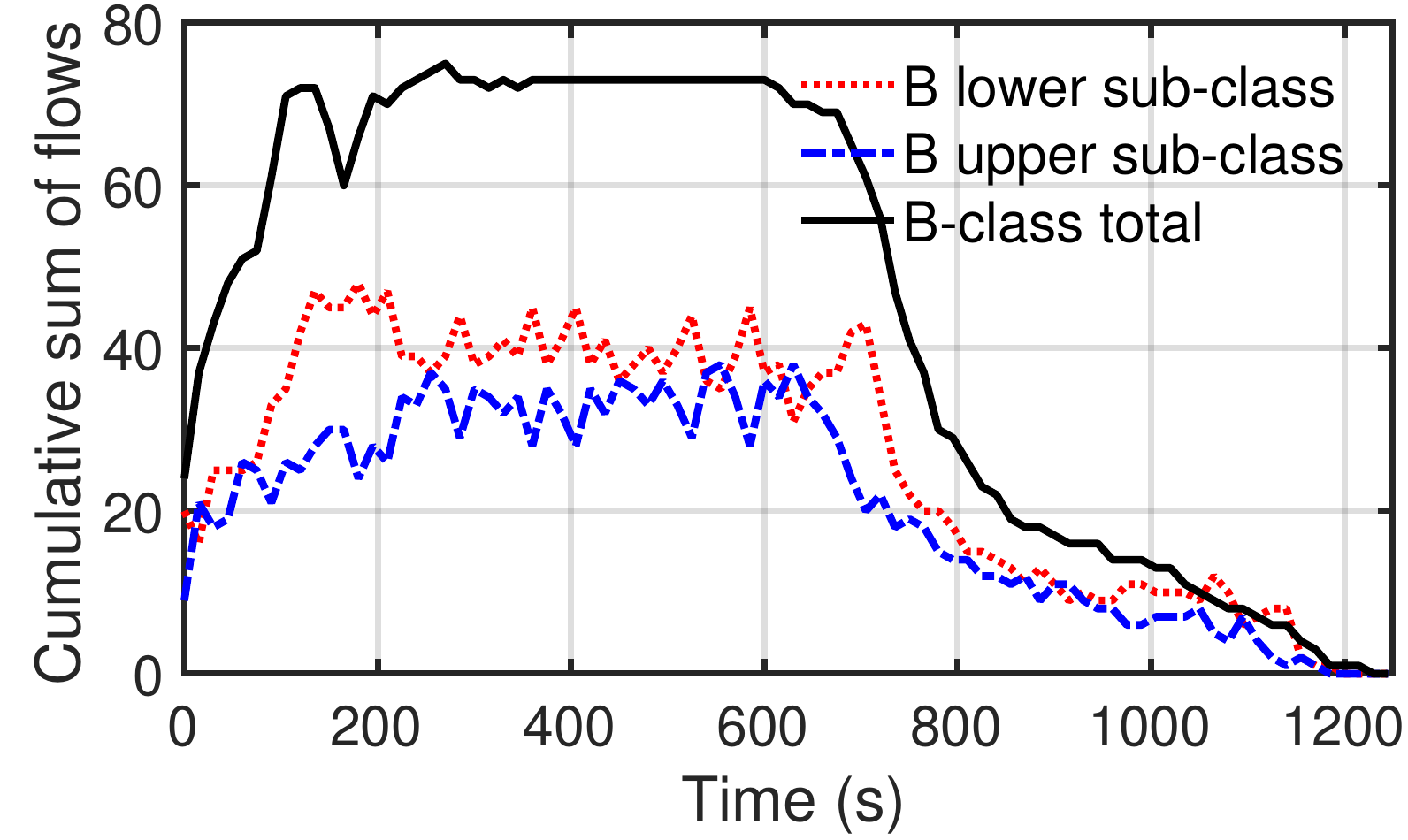}
    }
    \caption{DASH clients arrival-departure and service classes cumulative sum of active flows}
    \label{fig:dyn-sum}
\end{figure*}

\begin{figure*}
    \centering
    \subfigure[Bandwidth allocation for G-class]{ 
    \includegraphics[width=0.27\textwidth]{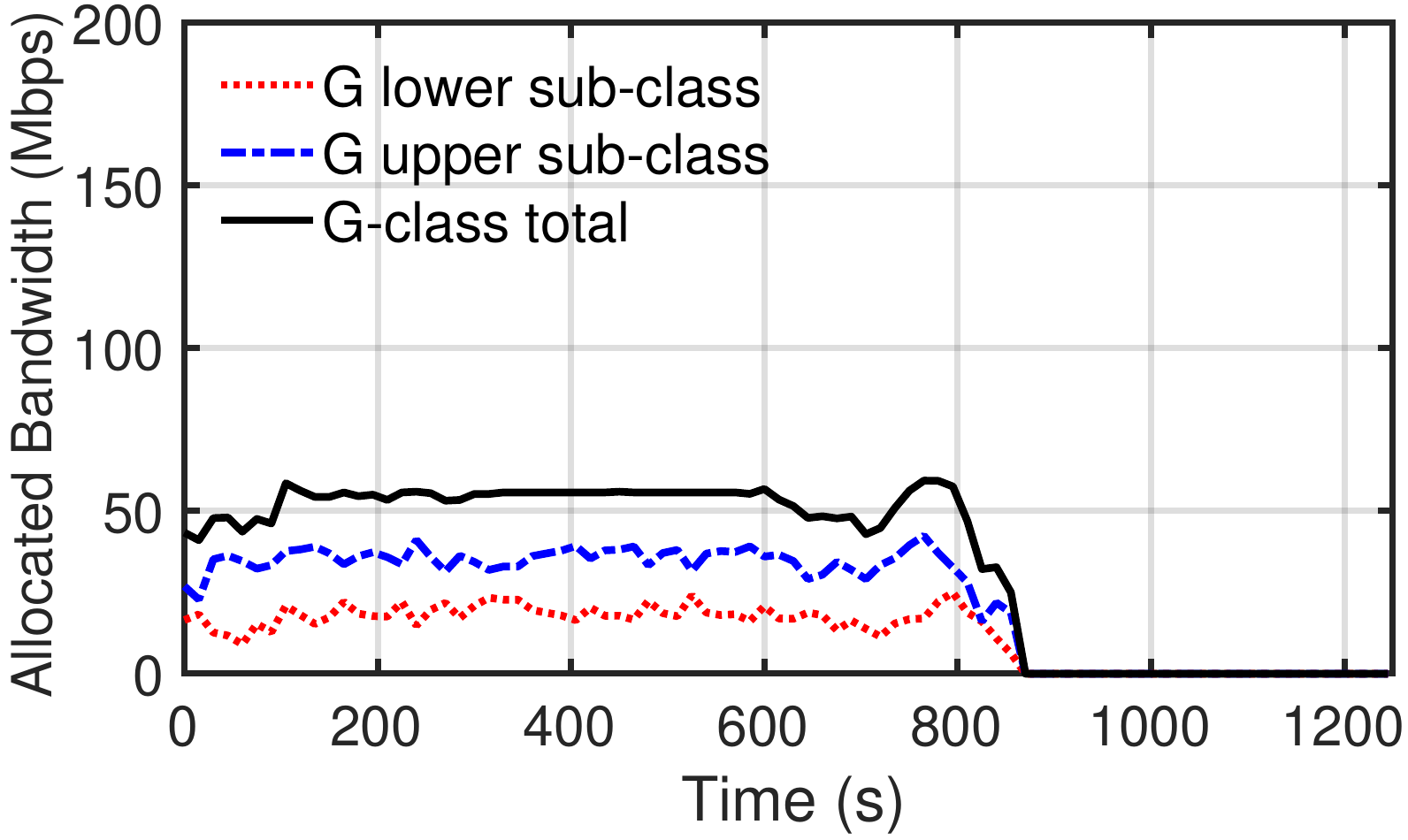}
    }
    \subfigure[Bandwidth allocation for S-class]{
    \includegraphics[width=0.27\textwidth]{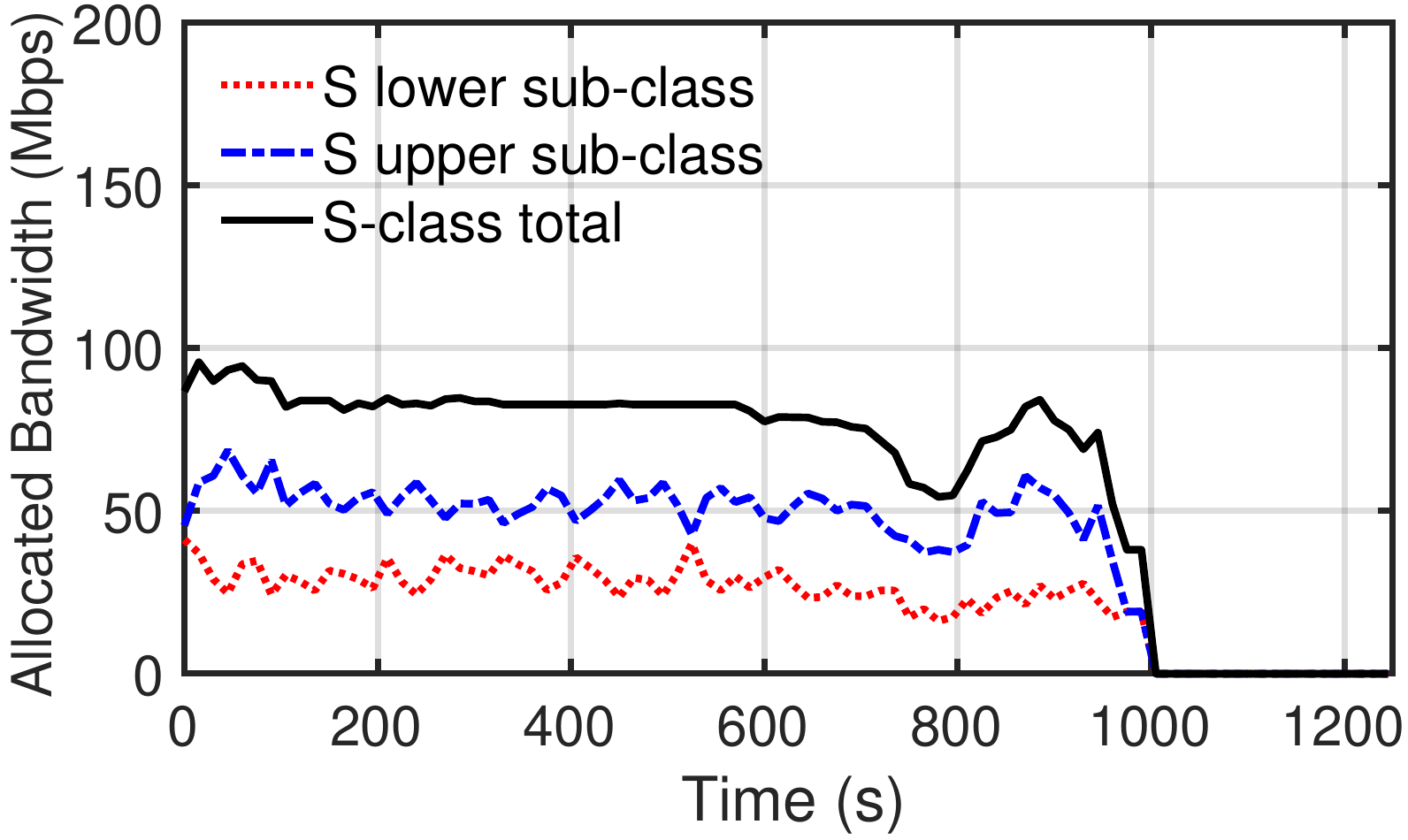}
    }
    \subfigure[Bandwidth allocation for B-class]{
    \includegraphics[width=0.27\textwidth]{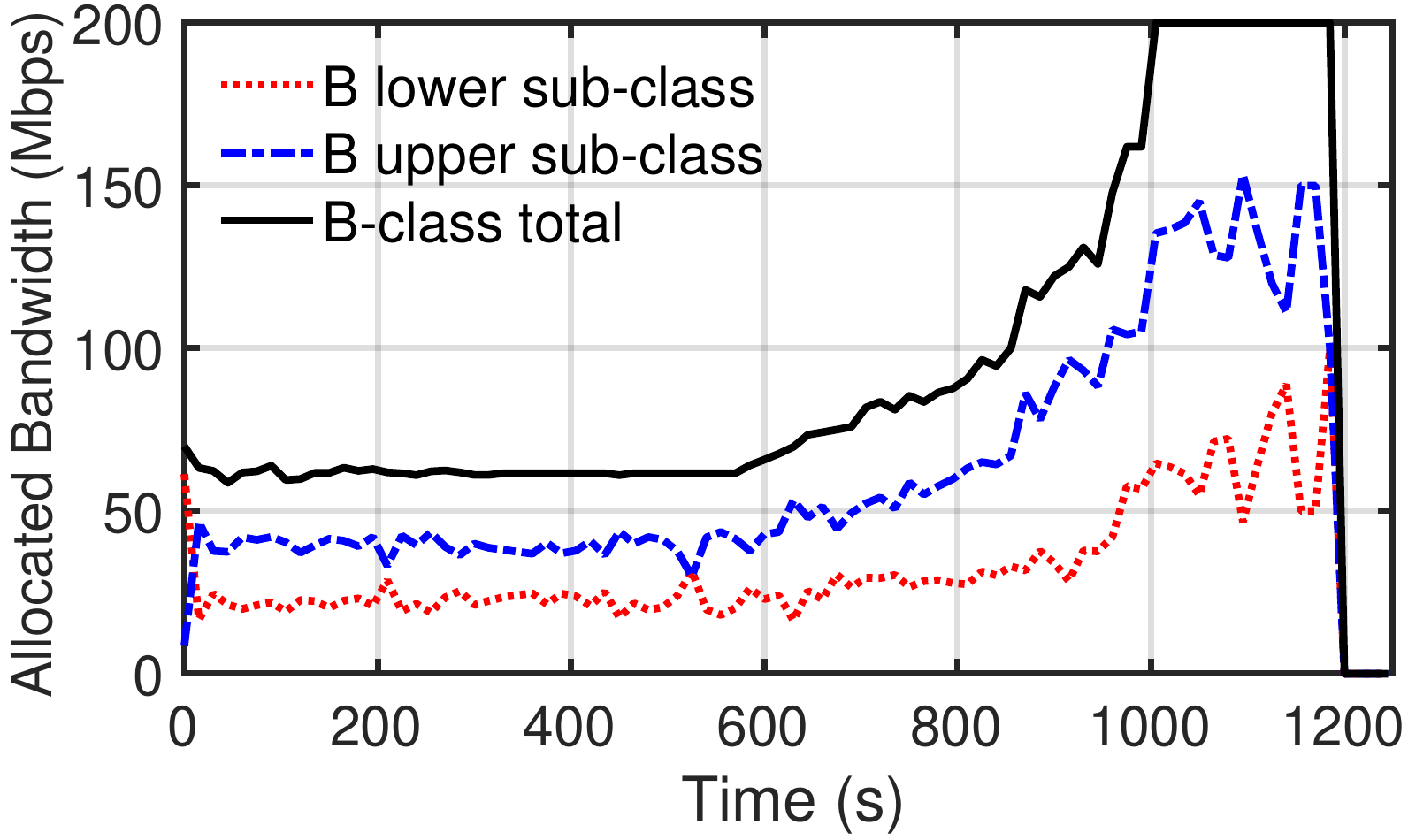}
    \label{B-class_bw}
    }
    \caption{Dynamic service classes bandwidth allocation}
    \label{fig:dyn-bw}
\end{figure*}
Overall, \diffperf is fairer than both CUBIC and BBR in terms of client's throughput, client's QoE, and provides the highest overall QoE.

\textbf{Scalability}: \diffperf operations on the Tofino switch are split across controller and dataplane. The controller collects aggregate real-time statistics of the active flows, performs the optimization, and reacts to data plane regularly. The dataplane tracks the number of  bytes transferred  by  the  active  flows. \diffperf does not impose high rate of sampling, which may lead to inaccurate statistics, especially when the DASH clients enter \texttt{OFF} period. Hence, the communication (between controller and dataplane) is only at the scale of seconds, and this works well for long running video flows over the Internet. This is also demonstrated by our experimental results.

\subsubsection{\underline{Scenario 3: The Dynamics of \diffperf}}
\label{s3}

Finally, to understand how \diffperf performs in real-world cases, we evaluate it in a dynamic scenario, where users from different service classes join and leave the network at different times.
In this set of experiments, we conduct the evaluation on OpenFlow network centralized by ODL controller, were 150 DASH clients share a $200$ Mbps bottleneck link; they have variable RTTs, with the ratio and distribution same as in the previous scenario. The arrivals of the DASH client requests follow the Poisson process with rate $\lambda= 1$ client/s. A client exits after the entire video (that lasts for $600$ seconds) is streamed.
The DASH clients subscribe to G, S, and B, service classes in the ratio 1:2:3. 
The weights of service classes are kept the same as before; i.e., G:S:B = 3:2:1. We set the values of $\alpha$, $\beta$ and $\gamma$ to $1$, $-0.25$, and $0.5$, respectively. At every interval of $\Delta_t$~=~15s, \diffperf  uses the last measured statistics, such as number of active flows and each flow's instantaneous throughput ($\delta$ = 0), to subsequently send command to the switch for regulating network flows in the next immediate time interval. OpenFlow switch updates \diffperf with flow statistics every 3s (i.e., the default sampling rate in Brocade ICX-6610 switch).
Figure~\ref{fig:dyn-sum}(a) shows the arrival and departure of DASH flows. The number of active flows in the two sub-classes, for each of the services classes, are depicted in Figures~\ref{fig:dyn-sum}(b),~(c), and~(d).  Although the video being streamed is of 600 seconds, observe that the G-class clients complete earlier than S-class and B-class; and this is true for both lower and upper sub-classes. 
Similarly, S-class flows finish earlier than B-class flows. 
We observe a sudden decrease in the active flows a few times (the dips on the curves); this
is not because the flow(s) actually leave the system, rather, due to the expiry of {\em idle timeout} of flows. 
When DASH client does not receive video segments packets, the timeout causes the flow to be deemed as an inactive flow. However, once the client resumes receiving data, \diffperf promptly counts it as an active flow.
Figure~\ref{fig:dyn-bw} plots the dynamic bandwidth allocation recommended by \diffperf, for each service class and the sub-classes within. The bandwidth allocated accounts for number of active flows in each service class and their achieved throughput, optimized via ($\beta$, $\gamma$) performance-aware mechanism. It also shows that \diffperf adapts quickly to the departure of flows (observe time period after 600 seconds), allocating the spare capacity to the remaining active flows. 

\section{Related Work}


\subsection{TCP Congestion Control}

Increase of network bandwidth also saw the emergence of `high-speed' TCP variants such as FAST~\cite{Fast-TCP-2006}, BIC~\cite{BIC-TCP-2004}, CUBIC~\cite{ha2008cubic} and BBR~\cite{BBR-Jacobson-2016} for transporting Internet traffic. 
Yet, TCP's inability to fairly share the bandwidth of flows with heterogeneous RTTs---a problem known to the community for around two decades~\cite{TCP-perf-survey-2000,Compound-TCP-2006,eval-TCP-HighSpeed-TON-2007}---still persists. 
As demonstrated by our experiments (and also other works, e.g.,~\cite{fairness-AINA-2008}) CUBIC exhibits such a behavior, and so does BBR~\cite{hock2017experimental,RTT-fairness-BBR-2017}. This unfairness in achieved throughput worsens when flows with different TCP congestion control mechanisms compete~\cite{CP-fairness-2019}. Another interesting observation from literature is that the relative performance degradation in throughput can be due to more than the single factor of RTT. In this context, we highlight that \diffperf is agnostic to the specific RTT of flows and other router specifications (e.g., buffer size) in performing optimization and enforcement of the computed optimized bandwidth; indeed \diffperf classifies and isolates flows of dissimilar characteristics solely based on tracking their
achieved throughput.

\subsection{Service Differentiation}

Service differentiation is at the core of network quality of service (QoS) provisioning to serve traffic from multiple classes over the network~\cite{diffserv,wang2001pricing,clark1994integrated,rao2012qos}. 
While {\em IntServ}~\cite{clark1994integrated} did not find adoption in the Internet, {\em DiffServ}~\cite{diffserv} inspired a body of work on providing differentiated services.
However, many of such solutions mandate a sophisticated scheduling with manual configuration of QoS knobs on a per service class basis. Instead, we choose a well-known utility function based framework which enables service operator to practically specify number of service classes and make good balance between the bandwidth share and performance.

In~\cite{RD}, authors proposed an approach for rate-delay (RD) differentiation by maintaining two queues at the router's output link. 
While the aspirations resemble \diffperf, it is still best-effort and does not promise any rate or loss guarantees. 
%
\cite{opentd} discussed a static service differentiation framework for ISP.
In short, class-based traffic control and service differentiation have been largely limited to theoretical analyses~\cite{RD,han2019utility,sung2009modeling, zou2018optimal}, and have not been experimented on hardware switches with real application traffic. 
~\diffperf's inter-class utility is general enough to make trade-offs among desirable performance metrics and operates dynamically based on active users and available bandwidth.

\subsection{Fair Queuing}
Fair queuing has been a topic for extensive research ~\cite{katabi2002congestion,hoeiland2016flowqueue,bennett1997hierarchical, hoeiland2016flowqueue, sharma2018approximating, hedayati2019multi, sivaraman2013no}. 
FQ-CoDel~\cite{hoeiland2016flowqueue}, a recent AQM discipline, offers good performance gain in achieving fairness among the flows, by classifying them into different buckets and serving them in a round-robin manner. However, large memory is extremely expensive or unavailable in data plane, hence it is practically infeasible to accommodate very large number of buckets for hashing large number of flows.
\diffperf optimizer performs simply by comparing flow's $z$ score with a pre-defined $\beta$ threshold, (i.e., classifier requires no training). Additionally, unlike AQM, \diffperf is not limited to specific congestion algorithm; it works on top of several interactive parameters such as buffer size, flow characteristics, and congestion algorithm. Also, \diffperf is portable, it can be packaged as virtual network function (VNF) over SmartNIC~\cite{SmartNICs}, to handle a presence of extremely heavy workload. 



\subsection{User Quality of Experience (QoE)}
In the context of video streaming, several studies proposed to improve user QoE~\cite{yadav,sdndash} or to achieve QoE fairness~\cite{georgopoulos2013towards}.
These approaches continuously attempt to improve the adaptive bitrate (ABR) algorithms in the DASH Reference Player at application layer, based on several performance metrics seen in the application. Our work differ from them in that we propose bottom-up optimization. \diffperf reacts to the interplay between several network's inherently coupled parameters by continuously improving affected traffic flows. This in large part improves the performance metrics (e.g., QoE fairness) at the application. 

\eat{
Additionally, \diffperf as service model is more aligned with the user requirements~\cite{dobrian2013understanding}, thus it increases network efficacy. 
The bottleneck buffer size plays a role in the QoE each client perceives, as confirmed in previous studies~\cite{hohlfeld2014qoe,spang2019buffer}.
Using throughput to estimate how flows are treated, \diffperf enforces fairness independent of the buffer size.
}

\section{Conclusion}

We propose \diffperf that leverages the rapid development in network softwarization and enables an agile and dynamic network bandwidth allocation at the AP vantage point. 
At a macroscopic level, \diffperf offers access providers new capabilities for performance guarantees 
by dynamically allocating bandwidth to service classes through which the trade-off between fairness and differentiation can be made.
At a microscopic level, \diffperf isolates and optimizes the affected flows as a result of interplay between  several  network’s  inherently  coupled parameters such as flow characteristics, buffer size, and congestion algorithm. We implemented two prototypes of \diffperf; one in ODL with OpenFlow, and the other on the programmable Tofino switch.
We evaluate \diffperf from an application perspective by evaluating it for MPEG-DASH video streaming. Our experiment results confirm \diffperf's capabilities of QoE provisioning, fairness, and optimization.

 \appendices

\section{Proofs of Theorems}

\textbf{Proof of Theorem \ref{thm:closed form}}: From optimization theory, our bandwidth allocation problem is a convex optimization problem.
By Karush-Kuhn-Tucker (KKT) conditions, it has a unique solution which satisfies
\begin{align*}
\begin{cases}
\displaystyle w_s\left(\frac{X_s}{n_s}\right)^{-\alpha} - u + u_G = 0 \ \ \text{and} \ \ u_s X_s =0,\ \forall s\in \mathcal{S},\\
\displaystyle u\left(\sum_{s\in\mathcal{S}} X_s - C\right) = 0
\end{cases}
\end{align*}
where \(u\) and \((u_s: s\in \mathcal{S})\) are KKT multipliers and satisfy \(u, u_s \ge 0\) for any \(s\in \mathcal{S}\).
By solving the above equations, we can derive that
\(X_s = \frac{n_s \sqrt[\alpha]{w_s}}{{\underset{s' \in \mathcal{S}}{\sum }n_{s'} \sqrt[\alpha]{w_{s'}} }} \ C, \forall s\in \mathcal{S}\).


\textbf{Proof of Theorem \ref{thm:monotonicity}}: 
By the definition of the set \(\mathcal{F}^L_s(\beta)\), we know that for any two thresholds \(\beta_1 < \beta_2\), \(\mathcal{F}^L_s(\beta_1) \subseteq \mathcal{F}^L_s(\beta_2)\). 
For any two flows \(\bar{f}\) and \({f}\) satisfying \(\bar{f}\in \mathcal{F}^L_s(\beta_1)\) and \({f}\in \mathcal{F}^L_s(\beta_2)\backslash \mathcal{F}^L_s(\beta_1)\), we have \(x_{\bar{f}} < x_{\tilde{f}}\) because \(z_{\bar{f}} < \beta_1 \le z_{\tilde{f}} < \beta_2\). 
Therefore, it satisfies 
$\frac{\sum_{f\in \mathcal{F}_s^L(\beta_1)}x_f}{|\mathcal{F}_s^L(\beta_1)|} \le \frac{\sum_{f\in \mathcal{F}_s^L(\beta_2)}x_f}{|\mathcal{F}_s^L(\beta_2)|}$,
i.e.,
the average achieved throughput of the flows within the lower sub-class~\(\mathcal{F}^L_s\) is non-decreasing in \(\beta\). By Eq.(\ref{eq:Eq 8}), when \(\beta = 0\), we have
\(\frac{\sum_{f\in \mathcal{F}_s^L(\beta)}x_f}{|\mathcal{F}_s^L(\beta)|} = \frac{\sum_{f\in \mathcal{F}_s^-}x_f}{|\mathcal{F}_s^-|} \le \frac{X_s^L(\beta)}{|\mathcal{F}_s^L(\beta)|}\)
because \(\mathcal{F}^L_s =\mathcal{F}^-_s\). In other words, the average achieved throughput of the flows within the lower sub-class~\(\mathcal{F}^L_s\) equals the average per-flow capacity re-allocated to~\(\mathcal{F}^L_s\).
Because \(|\mathcal{F}^L_s|\) is non-decreasing in \(\beta\), \(\gamma n_s/\left(n_s -|\mathcal{F}^L_s|\right)\) is non-decreasing in \(\beta\). By Eq.(\ref{eq:Eq 8}), the capacity allocated to the per-flow of the upper sub-class satisfies 
$\frac{X_s^H}{|\mathcal{F}^H_s|} = \frac{X_s - X^L_s}{|\mathcal{F}^H_s|}
=  \frac{\gamma n_s}{n_s -|\mathcal{F}^L_s|}\left(\frac{X_s}{n_s} - \frac{\sum_{f\in \mathcal{F}_s^-}x_f}{|\mathcal{F}_s^- |}\right)+\gamma \frac{\sum_{f\in \mathcal{F}_s^-}x_f}{|\mathcal{F}_s^- |} + (1-\gamma)\frac{X_s}{n_s} \ge \frac{X_s}{n_s}$.
Thus it is non-decreasing in \(\beta\) and no lower than \(X_s/n_s\).
 
\bibliographystyle{IEEEtran}
\bibliography{diffperf.bib}
\end{document}